\documentclass[12pt]{article}
\usepackage{epsfig}
\def\ket#1{\left| #1 \right\rangle}
\def\vev#1{\left\langle #1 \right\rangle}
\begin{document}
\baselineskip=15.5pt
\renewcommand{\theequation}{\arabic{section}.\arabic{equation}}
\pagestyle{plain}
\setcounter{page}{1}
\def\appendix{
\par
\setcounter{section}{0}
\setcounter{subsection}{0}
\def\thesection{\Alph{section}}}
\begin{titlepage}

\rightline{\tt hep-th/0310264}
\rightline{\small{\tt MIT-CTP-3429}}
\rightline{\small{\tt CALT-68-2460}}

\begin{center}

\vskip 1.5cm

{\LARGE {Some exact computations}}

\vskip 0.5cm

{\LARGE {on the twisted butterfly state}}  

\vskip 0.5cm

{\LARGE {in string field theory}}  

\vskip 2.0cm

{\large {Yuji Okawa}}

\vskip 0.5cm

{\it {Center for Theoretical Physics, Room 6-304}}

\smallskip

{\it {Massachusetts Institute of Technology}}

\smallskip

{\it {Cambridge, MA 02139, USA}}

\smallskip

okawa@lns.mit.edu

\vskip 2.0cm

{\bf Abstract}
\end{center}

\noindent
The twisted butterfly state solves the equation of motion
of vacuum string field theory in the singular limit.
The finiteness of the energy density of the solution
is an important issue,
but possible conformal anomaly resulting from the twisting
has prevented us from addressing this problem.
We present a description of the twisted regulated butterfly state
in terms of a conformal field theory with a vanishing central charge
which consists of the ordinary $bc$ ghosts
and a matter system with $c=26$.
Various quantities relevant to vacuum string field theory
are computed exactly using this description.
We find that the energy density of the solution
can be finite in the limit,
but the finiteness depends on the subleading structure
of vacuum string field theory.
We further argue, contrary to our previous expectation,
that contributions from subleading terms in the kinetic term
to the energy density can be of the same order
as the contribution from the leading term which consists of
the midpoint ghost insertion.

\end{titlepage}

\newpage

\section{Introduction}
\setcounter{equation}{0}

Vacuum string field theory
\cite{Rastelli:2000hv, Rastelli:2001jb, Rastelli:2001uv}
is a conjectured form of
Witten's cubic open string field theory \cite{Witten:1985cc}
expanded around the tachyon vacuum.\footnote
{
For a recent review, see \cite{Taylor:2003gn}.
}
The action of vacuum string field theory
is given by replacing the BRST operator
in Witten's string field theory
with a different operator ${\cal Q}$ which is conjectured to be
made purely of ghost fields \cite{Rastelli:2000hv}.
The equation of motion of vacuum string field theory,
\begin{equation}
  {\cal Q} \ket{\Psi} + \ket{\Psi \ast \Psi} = 0,
\end{equation}
then factorizes into the matter and ghost sectors,
and the matter part of the equation can be solved
by the matter sector of a star-algebra projector
\cite{Rastelli:2001jb, Kostelecky:2000hz}.
Using the conformal field theory (CFT) formulation
of string field theory \cite{LeClair:1988sp, LeClair:1988sj},
we can construct such a star-algebra projector
for any given consistent open-string boundary condition
\cite{Rastelli:2001vb}.
The resulting solution is conjectured to describe the D-brane
corresponding to the open-string boundary condition
\cite{Rastelli:2001vb, Mukhopadhyay:2001ey}.
Based on this description of D-branes
in vacuum string field theory,
it has been shown that ratios of D-brane tensions
\cite{Rastelli:2001vb, Okuyama:2002tw},
the open-string mass spectrum on any D-brane \cite{Okawa:2002pd},
and the absolute value of the D25-brane tension \cite{Okawa:2002pd}
can be reproduced correctly.

One important assumption in deriving these results
is the existence of a solution to the ghost part of the equation.
Moreover, in the derivation of the absolute value of the D25-brane
tension in \cite{Okawa:2002pd},
the energy density of the full solution
which consists of both matter and ghost sectors
was related to the on-shell three-tachyon coupling constant
on a D25-brane so that the energy density must be finite
in order to have a finite string coupling constant.

A solution to the equation of motion of vacuum string field theory
was first found by Hata and Kawano \cite{Hata:2001sq}
in the operator formulation
\cite{Gross:1986ia, Gross:1986fk, Ohta:wn,
Cremmer:if, Samuel:1986wp}
with a particular choice of the kinetic operator ${\cal Q}$.
It turned out later that their solution is the sliver state
in the twisted ghost CFT \cite{Gaiotto:2001ji, Okuda:2002fj},
and their kinetic operator is a $c$-ghost insertion
at the open-string midpoint \cite{Okuyama:2002yr}.
It was further shown in \cite{Gaiotto:2001ji} that
any star-algebra projector in the twisted ghost CFT
solves the equation of motion of vacuum string field theory
with ${\cal Q}$ being the midpoint $c$-ghost insertion.

However, what has been shown in solving the equation of motion
of vacuum string field theory is the proportionality between
${\cal Q} \ket{\Psi}$ and $\ket{\Psi \ast \Psi}$.
As long as the proportionality constant is finite,
the equation of motion can be satisfied
by a finite rescaling of the solution.
However, if it is infinite or vanishing,
the normalization of the solution becomes singular
so that regularization would be necessary to make it well-defined.
In fact, the normalization of the Hata-Kawano solution
seems singular \cite{Hata:2001sq, Okuyama:2002yr}.
Even if the solution itself is well-defined,
we may encounter singularities when we compute
$\vev{ \Psi | {\cal Q} | \Psi }$
and $\vev{ \Psi | \Psi \ast \Psi }$
to evaluate the energy density of the solution.
In fact, various solutions of string field theory
based on the identity string field
\cite{Horowitz:dt, Takahashi:2001pp, Kishimoto:2001de,
Kluson:2002ex, Takahashi:2002ez, Kluson:2002hr, Kishimoto:2002xi,
Kluson:2002gu, Kluson:2002te, Kluson:2002av, Takahashi:2003pp,
Kluson:2003xu, Takahashi:2003xe}
suffer from the notorious singularity coming from the inner product
of the identity string field with itself.
Furthermore, even if the quantities
$\vev{ \Psi | {\cal Q} | \Psi }$
and $\vev{ \Psi | \Psi \ast \Psi }$
can be made well-defined,
it is still a nontrivial question
whether or not the equation of motion is satisfied
when it is contracted with the solution itself,
namely, whether or not
$\vev{ \Psi | {\cal Q} | \Psi } + \vev{ \Psi | \Psi \ast \Psi } = 0$
holds.
We therefore recognize that there are these nontrivial steps
to establish the existence of a solution with a finite energy density.

One technical difficulty which has prevented us from addressing
these questions in the CFT approach is possible conformal anomaly
coming from the twisting.
In this paper, we present
a description of a class of twisted surface states in terms of
the system of the ordinary $bc$ ghosts and a matter CFT with $c=26$
to overcome this difficulty.
Since the total central charge vanishes,
no conformal anomaly arises when we make a conformal transformation
in the process of gluing string fields.
This is important because
the generalized gluing and resmoothing theorem \cite{LeClair:1988sj}
holds only when the total central charge vanishes
as was emphasized in \cite{Asakawa:1998dv}.
We in particular study the twisted regulated butterfly state
\cite{Gaiotto:2001ji, Schnabl:2002ff, Gaiotto:2002kf}
in detail, and compute various quantities involving this state
exactly.

We find that the proportionality constant between
${\cal Q} \ket{\Psi}$ and $\ket{\Psi \ast \Psi}$ is in fact singular
in the case of the twisted butterfly state.
However, the energy density of the solution can be finite
in the limit by appropriately scaling the normalization of
the twisted regulated butterfly state
and the kinetic operator ${\cal Q}$.

This is good news for the vacuum string field theory conjecture.
However, there is another subtlety.
The conjectured kinetic operator of vacuum string field theory
consisting of the midpoint $c$-ghost insertion
with a divergent coefficient requires regularization
\cite{Gaiotto:2001ji, Schnabl:2002ff, Gross:2001yk}
and, as we will show in this paper,
the subleading structure of the kinetic operator
is necessary in order
for vacuum string field theory to have a parameter
corresponding to the string coupling constant.
The question is then whether or not
details of the regularization contribute to the physics
in the limit.
Apparently, the midpoint ghost insertion dominates
in the kinetic operator
and the twisted butterfly state provides a formal solution
in the limit.
However, we find that
subleading terms of the kinetic term
can contribute to the energy density
at the same order as the leading term.
This is not an immediate problem of vacuum string field theory,
but the subleading structure may ruin the factorization of
the matter and ghost sectors at the leading order.
In this paper, we present the first quantitative approach
to this issue.

The organization of the paper is as follows.
We provide the description
of the twisted regulated butterfly state
in terms of the ordinary $bc$ ghosts and a matter CFT
with $c=26$ in Section 2.
We then present exact computations of various quantities
involving the twisted regulated butterfly state
in Section 3.
The issues raised in the introduction are discussed
in Section 4.
Section 5 is devoted to conclusion and discussion.
Our conventions and terminology on the CFT formulation
of string field theory are summarized in Appendix A.
Some details of computations in Sections 2 and 3
are given in Appendices B and C.

\section{Twisted regulated butterfly state
in terms of the ordinary ghost CFT}
\setcounter{equation}{0}

\subsection{Regulated butterfly state}

The butterfly state is a star-algebra projector
originally found by the level-truncation analysis
of vacuum string field theory in \cite{Gaiotto:2001ji},
and its properties were further studied
in \cite{Schnabl:2002ff, Gaiotto:2002kf}.
The regulated butterfly state $\ket{B_t}$
introduced in \cite{Schnabl:2002ff, Gaiotto:2002kf}
is a regularization of the butterfly state
parameterized by $t$ in the range $0 \le t < 1$.
It is defined by
\begin{equation}
  \vev{ \phi | B_t } = \vev{ f_t \circ \phi (0) }
\end{equation}
for any state $\ket{\phi}$ in the Fock space, where
\begin{equation}
  f_t (\xi) = \frac{\xi}{\sqrt{1 + t^2 \xi^2}}.
\end{equation}
All CFT correlation functions in this paper are evaluated
on an upper-half plane, and we use the doubling trick.

The butterfly state $\ket{B}$ is given
by the regulated butterfly state in the limit $t \to 1$.
It is a singular state like other star-algebra projectors
such as the sliver state \cite{Moore:2001fg}.
The singularity can be seen, for example, by the fact that
the open-string midpoint $f_t (i)$ reaches the boundary
in the limit $t \to 1$.
However, an inner product of the regulated butterfly state
with a state $\ket{\phi}$ in the Fock space is well-defined
even in the limit $t \to 1$ and is given by
\begin{equation}
  \vev{ \phi | B } = \lim_{t \to 1} \vev{ \phi | B_t }
  = \vev{ f_B \circ \phi (0) },
\end{equation}
where
\begin{equation}
  f_B (\xi) = \frac{\xi}{\sqrt{1 + \xi^2}}.
\end{equation}
In the opposite limit $t \to 0$,
the regulated butterfly state reduces to
the $SL(2,R)$-invariant vacuum $\ket{0}$.

We can use different conformal transformations
to represent the same surface state.
For example, the regulated butterfly state
can also be represented as
\begin{equation}
  \vev{ \phi | B_t } = \vev{ h_t \circ \phi (0) }
\end{equation}
for any state $\ket{\phi}$ in the Fock space,
where $h_t (\xi)$ is a conformal transformation
with a parameter $p$:
\begin{equation}
  h_t (\xi) = \frac{\xi}{\xi + p \sqrt{1 + t^2 \xi^2}}.
\label{h_t}
\end{equation}
The conformal transformation $h_t (\xi)$ is related to
$f_t (\xi)$ by an $SL(2,R)$ transformation $z/(z+p)$ 
which maps the infinity to $1$.
The conformal transformations
with different values of $p$ for (\ref{h_t})
are all equivalent and define the same state.
This representation will be useful when there is an operator
insertion at the infinity
in the representation in terms of $f_t (\xi)$.
In this representation, the inner product $\vev{ \phi | B }$
in the butterfly limit is given by
\begin{equation}
  \vev{ \phi | B } = \vev{ h_B \circ \phi (0) },
\end{equation}
where
\begin{equation}
  h_B (\xi) = \frac{\xi}{\xi + p \sqrt{1 + \xi^2}}.
\end{equation}

The regulated butterfly state has a simple representation
in terms of a single Virasoro generator
\cite{Schnabl:2002ff, Gaiotto:2002kf}.
It is given by
\begin{equation}
  \ket{B_t} = \exp \left( -\frac{t^2}{2} L_{-2} \right) \ket{0}.
\end{equation}

\subsection{Twisted regulated butterfly state}

The twisted ghost CFT introduced in \cite{Gaiotto:2001ji}
in the context of vacuum string field theory
is defined by changing the energy-momentum tensor
in the strip coordinates as
\begin{equation}
  T'(w) = T(w) - \partial \, j_g (w), \quad
  \widetilde{T}'(\bar{w}) = \widetilde{T}(\bar{w})
  - \bar{\partial} \, \widetilde{\jmath}_g (\bar{w}).
\end{equation}
A twisted surface state is defined
by a correlation function in the twisted ghost CFT.
The twisted regulated butterfly state $\ket{B'_t}$
is defined by
\begin{equation}
  \vev{ \phi | B'_t } = \vev{ h_t \circ \phi' (0) }'
\label{B'_t-twisted}
\end{equation}
for any state $\ket{\phi}$ in the Fock space,
and the twisted butterfly state $\ket{B'}$
is given by the singular limit of $\ket{B'_t}$:
\begin{equation}
  \vev{ \phi | B' } = \lim_{t \to 1} \vev{ \phi | B'_t }
  = \vev{ h_B \circ \phi' (0) }'.
\end{equation}
The prime on the correlation functions denotes
that they are evaluated in the twisted CFT.
Note that the state-operator correspondence is modified
because the twisting changes the conformal properties of
the ghost fields.
For example, the state $c_1 \ket{0}$
corresponds to the operator $c(0)$ in the ordinary $bc$ CFT,
but it corresponds to the identity operator
in the twisted CFT.
In other words, the state $c_1 \ket{0}$ corresponds to
the $SL(2,R)$-invariant vacuum $\ket{0'}$
in the twisted CFT.
We denoted the operator in the twisted CFT
corresponding to $\ket{\phi}$ by $\phi'(0)$.
In the operator formalism, the twisted regulated butterfly state
is represented as
\begin{equation}
  \ket{B'_t} = \exp \left( -\frac{t^2}{2} L'_{-2} \right) \ket{0'}
  = \exp \left( -\frac{t^2}{2} L'_{-2} \right) c_1 \ket{0},
\label{B'_t-operator}
\end{equation}
where $L'_{-2}$ is the Virasoro generator in the twisted CFT.

\subsection{Twisted regulated butterfly state
in terms of the ordinary ghost CFT}

Because of the twisting, the total central charge of
the system consisting of the twisted ghost CFT
and a matter CFT with $c=26$ no longer vanishes.
Therefore, we have to deal with conformal anomaly.
It would be useful if we can represent twisted surface states
in terms of correlation functions of CFT
with a vanishing central charge.
Let us try to find such a representation
of the twisted regulated butterfly state $\ket{B'_t}$.

The twisting corresponds to a different coupling
to the world-sheet curvature in the language of bosonization.
The world-sheet curvature vanishes in the strip coordinates
so that we can easily find the relation between
representations in the ordinary and twisted CFT's
for a state in the Fock space.
For example, the state $c_1 \ket{0}$
corresponds to the insertion of $c(0)$ in the ordinary $bc$ CFT
and to the insertion of the identity operator in the twisted CFT.

The regulated butterfly state is not
a state in the Fock space, but it takes a simple form
in the strip coordinates $(\tau, \sigma)$
where a state in the Fock space
is represented as a path integral
over the semi-infinite strip
$-\infty < \tau \le 0$, $0 \le \sigma \le \pi$
with a wave function in the infinite past $\tau = -\infty$.
The regulated butterfly state is represented
as a path integral over this region with a slit from
$(\tau, \sigma) = (-\infty, \pi/2)$ to
$(\tau, \sigma) = (\ln t, \pi/2)$.
The boundary of the surface runs from $(\tau, \sigma) = (0, \pi)$
to $(\tau, \sigma) = (0, 0)$ as follows:
$(0, \pi) \to (-\infty, \pi),
(-\infty, \pi/2) \to (\ln t, \pi/2) \to (-\infty, \pi/2),
(-\infty, 0) \to (0,0)$.
When $t=0$, the slit vanishes
and the state reduces to the $SL(2,R)$-invariant vacuum $\ket{0}$.
In the limit $t \to 1$, the boundary reaches the open-string midpoint
$(\tau, \sigma) = (0, \pi/2)$, and the wave function at $\tau=0$
factorizes into those of the left and right half-strings.
The world-sheet curvature vanishes in the interior of this region,
and it also vanishes on the boundary except for a point
$(\tau, \sigma) = (\ln t, \pi/2)$.
The wave functions at $\tau = -\infty$ in the regions
$0 \le \sigma \le \pi/2$ and $\pi/2 \le \sigma \le \pi$
should correspond to insertions of $c$ ghosts when they are mapped
to a point just as in the case of
the $SL(2,R)$-invariant vacuum $\ket{0}$.
Note that the boundary changes its direction
by the amount of $\pi$
when it goes to $\tau = -\infty$ and comes back.
At $(\tau, \sigma) = (\ln t, \pi/2)$,
the change in the direction is $-\pi$.
Therefore, the $b$ ghost should be inserted
in the ordinary CFT
when this point is mapped to a point
where the boundary does not bend.

To summarize, the twisted regulated butterfly state $\ket{B'_t}$,
which is a surface state without any operator insertions
in the twisted CFT, should be described as the same surface state
with two $c$-ghost and one $b$-ghost insertions
in the ordinary $bc$ CFT.
If we use the conformal transformation $f_t (\xi)$ in the definition,
the insertion points of the $c$ ghosts are $-1/t$ and $1/t$,
and that of the $b$ ghost is the infinity, which is outside
of the patch.
If we use the conformal transformation $h_t (\xi)$ instead,
all these three points are simultaneously finite.
The twisted regulated butterfly state $\ket{B'_t}$
is then represented as follows:
\begin{equation}
  \vev{ \phi | B'_t } = -\frac{t}{2} (p^2 t^2 -1)^2
  \vev{ c \left( \frac{1}{1+pt} \right) b (1) \,
        c \left( \frac{1}{1-pt} \right) h_t \circ \phi (0) }
\label{B'_t}
\end{equation}
for any state $\ket{\phi}$ in the Fock space.
The $p$-dependence of the normalization factor in (\ref{B'_t})
is determined by the covariance under the $SL(2,R)$ conformal
transformation $pz/(pz+q(1-z))$
which changes the parameter $p$ to $q$.
The normalization is then fixed by the condition
$\vev{B'_t | c_0 c_1 | 0}_{density} =1$
for any $t$ in $0 \le t < 1$,
where the subscript $density$ denotes that
the inner product has been divided by the space-time volume.
This normalization evidently coincides with that
in the representation (\ref{B'_t-operator}).
The definitions with different values of $p$
are all equivalent, and the inner product $\vev{ \phi | B'_t }$
is independent of $p$.
The twisted regulated butterfly state
is now represented by a correlation function
in the system of the ordinary $bc$ ghosts
and a matter CFT with $c=26$,
which is free from conformal anomaly.
We will present an explicit proof of the equivalence
between this representation
and the definition (\ref{B'_t-twisted}) in the next subsection.

As a consistency check, let us consider the limit $t \to 0$.
The two $c$ ghosts approach the $b$ ghost in the limit
so that these three operators can be replaced by the leading term
in the operator product expansion (OPE).
Since
\begin{equation}
  c \left( \frac{1}{1+pt} \right) b (1) \,
  c \left( \frac{1}{1-pt} \right) = -\frac{2}{pt} \, c(1) + O(1),
\end{equation}
the inner product $\vev{ \phi | B'_t }$ becomes
\begin{equation}
  \vev{ \phi | B'_t } \to \frac{1}{p}
  \vev{ c(1) \, h_{t=0} \circ \phi (0) }
\end{equation}
in the limit $t \to 0$.
It can be easily verified that
this coincides with $\vev{ \phi | c_1 | 0}$.
Therefore, the representation of $\ket{B'_t}$ correctly
reduces to that of $c_1 \ket{0}$ in the limit $t \to 0$.

There is another useful representation of the twisted
regulated butterfly state $\ket{B'_t}$.
Let us make an inversion $I(z)=-1/z$
to the conformal transformation $f_t (\xi)$:
\begin{equation}
  I \circ f_t (\xi) = -\frac{\sqrt{1 + t^2 \xi^2}}{\xi}.
\end{equation}
Using this conformal transformation,
the twisted regulated butterfly state is represented as
\begin{equation}
  \vev{ \phi | B'_t } = -\frac{t}{2}
  \vev{ c(-t) \, b (0) \, c(t) \, I \circ f_t \circ \phi (0) }.
\label{B'_t-another}
\end{equation}
The origin $\xi=0$ is mapped to the infinity by $I \circ f_t (\xi)$,
but the three ghost-insertion points are finite.
The coordinate $z'=I \circ f_t (\xi)$ is related to
the coordinate $z=h_t (\xi)$ by
\begin{equation}
  z' = \frac{z-1}{p z}, \quad z = \frac{1}{1 - p z'}.
\end{equation}

The twisted regulated butterfly
state satisfies the Siegel gauge condition
$\vev{\phi | b_0 | B'_t}=0$ for an arbitrary $t$.
This is obvious in the representation
(\ref{B'_t-operator}).
It can also be easily seen
in the representation (\ref{B'_t-another})
in the following way.
If we define $z = I \circ f_t (\xi)$, $b_0$ can be expressed as
\begin{equation}
  b_0 = \oint \frac{d \xi}{2 \pi i} \, \xi \, b(\xi)
  = \oint \frac{dz}{2 \pi i} \, \frac{t^2-z^2}{z} \, b(z),
\label{b_0}
\end{equation}
where the contour of the $\xi$-integral encircles the origin
counterclockwise and that of the $z$-integral encircles
$-t$, the origin, and $t$ counterclockwise.
The pole in the OPE between $b(z)$ and $c(\pm t)$ is canceled
by the zero in (\ref{b_0}), and the pole in (\ref{b_0}) at $z=0$
is canceled by the zero in the OPE between $b(z)$ and $b(0)$.
Therefore, $\vev{\phi | b_0 | B'_t}=0$ for an arbitrary $t$
in the representation (\ref{B'_t-another}).

\subsection{Proof of the equivalence}

In this subsection, we will verify our representation (\ref{B'_t})
of the twisted regulated butterfly state
by providing an explicit proof of
the equivalence
between $\ket{B'_t}$ and $| \widetilde{B}'_t \rangle$
which are defined by
\begin{equation}
  \vev{ B'_t | \phi } = -\frac{t}{2} (p^2 t^2 -1)^2
  \vev{ c \left( \frac{1}{1+pt} \right) b (1) \,
        c \left( \frac{1}{1-pt} \right) h_t \circ \phi (0) }
\end{equation}
and
\begin{equation}
  \langle \widetilde{B}'_t | \phi \rangle
  = \vev{ h_t \circ \phi' (0) }',
\end{equation}
for any state $\ket{\phi}$ in the Fock space, respectively.\footnote
{
The ghost number of $\ket{B'_t}$ is one
so that the ghost number of $\ket{\phi}$ must be two
in order to have a nonvanishing inner product.
Therefore, $\vev{ B'_t | \phi }=\vev{ \phi | B'_t }$.
Similarly, $\langle \widetilde{B}'_t | \phi \rangle
= \langle \phi | \widetilde{B}'_t \rangle$.
}

Let us begin with the case
where $\ket{\phi} = c_{-n} b_{-m} c_0 c_1 \ket{0}$, and define
\begin{equation}
  M_{nm} \equiv \vev{ B'_t | c_{-n} b_{-m} c_0 c_1 | 0 }, \quad
  \widetilde{M}_{nm} \equiv
  \langle \widetilde{B}'_t | c_{-n} b_{-m} c_0 c_1 | 0 \rangle,
\end{equation}
and
\begin{equation}
  M (z,w) \equiv \vev{ B'_t | c(w) b(z) c_0 c_1 | 0 }, \quad
  \widetilde{M} (z,w) \equiv
  \langle \widetilde{B}'_t | c'(w) b'(z) c_0 | 0' \rangle.
\end{equation}
Note that $\ket{0'} = c_1 \ket{0}$.
Since the modes $c_n$ and $b_n$ are related to the ordinary
and twisted ghosts as
\begin{eqnarray}
  && c_n = \oint \frac{dw}{2 \pi i} \, w^{n-2} \, c(w)
      = \oint \frac{dw}{2 \pi i} \, w^{n-1} \, c'(w),
\label{c_n}
\\
  && b_m = \oint \frac{dz}{2 \pi i} \, z^{m+1} \, b(z)
      = \oint \frac{dz}{2 \pi i} \, z^{m} \, b'(z),
\end{eqnarray}
respectively, we have
\begin{equation}
  M_{nm} = \oint \frac{dw}{2 \pi i} \, \frac{1}{w^{n+2}}
           \oint \frac{dz}{2 \pi i} \, \frac{1}{z^{m-1}} \, M (z,w),
\end{equation}
and
\begin{equation}
  \widetilde{M}_{nm} = \oint \frac{dw}{2 \pi i} \, \frac{1}{w^{n+1}}
  \oint \frac{dz}{2 \pi i} \, \frac{1}{z^{m}} \, \widetilde{M} (z,w).
\end{equation}
Therefore, if $M(z,w)$ is $w/z$ times $\widetilde{M} (z,w)$,
the two inner products $M_{nm}$ and $\widetilde{M}_{nm}$ coincide.

Let us compute $\widetilde{M} (z,w)$ and $M (z,w)$.
In the twisted ghost CFT, the state $\ket{0'}$ corresponds to
the identity operator, and $c_0 \ket{0'}$ corresponds to
the operator $c'(0)$.
Therefore, $\widetilde{M} (z,w)$ is given by
\begin{eqnarray}
  \widetilde{M} (z,w) &=& \vev{ h_t \circ c'(w) \,
    h_t \circ b'(z) \, h_t \circ c'(0) }'
\nonumber \\
  &=& \frac{d h_t (z)}{dz}
      \vev{ c'( h_t (w) ) \, b'( h_t (z) ) \, c'(0)}'
\nonumber \\
  &=& \frac{d h_t (z)}{dz}
      \frac{1}{h_t (w) - h_t (z)}
      \frac{h_t (w)}{h_t (z)}.
\label{M-tilde}
\end{eqnarray}
Note that the conformal dimensions of $b'$ and $c'$ are
$1$ and $0$, respectively.
In the ordinary ghost CFT, the state $c_1 \ket{0}$ corresponds to
the operator $c(0)$, and $c_0 c_1 \ket{0}$ corresponds to
$- c \partial c (0)$, which is a primary field
and its conformal dimension is $-1$.
Therefore, $M(z,w)$ is given by
\begin{eqnarray}
  && M(z,w) \nonumber \\
  && = \frac{t}{2} (p^2 t^2 -1)^2
  \vev{ c \left( \frac{1}{1+pt} \right) b (1) \,
        c \left( \frac{1}{1-pt} \right) h_t \circ c(w) \,
        h_t \circ b(z) \, h_t \circ c \partial c(0) }
\nonumber \\
  && = \frac{p t}{2} (p^2 t^2 -1)^2
    \left( \frac{d h_t (z)}{dz} \right)^2
    \left( \frac{d h_t (w)}{dw} \right)^{-1}
\nonumber \\
  && \qquad \times
     \vev{ c \left( \frac{1}{1+pt} \right) b (1) \,
           c \left( \frac{1}{1-pt} \right)  c( h_t(w) ) \,
           b( h_t(z) ) \, c \partial c(0) }
\nonumber \\
  && = \left( \frac{d h_t (z)}{dz} \right)^2
     \left( \frac{d h_t (w)}{dw} \right)^{-1}
     \frac{1}{ h_t(w) - h_t(z) }
\nonumber \\
  && \qquad \times
     \left( \frac{1}{1+pt} - h_t(w) \right)
     \left( \frac{1}{1-pt} - h_t(w) \right)
     \frac{ h_t(w)^2 }{1- h_t(w)}
\nonumber \\
  && \qquad \times
     \left[ \left( \frac{1}{1+pt} - h_t(z) \right)
     \left( \frac{1}{1-pt} - h_t(z) \right)
     \frac{ h_t(z)^2 }{1- h_t(z)} \right]^{-1}.
\label{M}
\end{eqnarray}
This expression for $M(z,w)$ looks very different from
(\ref{M-tilde}) for $\widetilde{M} (z,w)$.
However, since
\begin{equation}
  \frac{d h_t (z)}{dz}
  = \frac{p^2 t^2 -1}{z}
    \frac{ h_t(z) }{h_t(z)-1}
    \left( h_t(z) - \frac{1}{1+pt} \right)
    \left( h_t(z) - \frac{1}{1-pt} \right),
\label{f^p-identity}
\end{equation}
the expression (\ref{M}) can be simplified to give
\begin{equation}
  M (z,w) = \frac{w}{z} \frac{d h_t (z)}{dz}
            \frac{1}{h_t (w) - h_t (z)}
            \frac{h_t (w)}{h_t (z)}
          = \frac{w}{z} \widetilde{M} (z,w).
\end{equation}
Thus we have shown that $M_{nm}=\widetilde{M}_{nm}$.

It is straightforward to generalize the proof for
an arbitrary $\ket{\phi}$. Let us define
\begin{eqnarray}
  M_{n_1 m_1 n_2 m_2 \cdots n_k m_k} &\equiv&
  \vev{ B'_t | c_{-n_1} b_{-m_1} c_{-n_2} b_{-m_2} \cdots
        c_{-n_k} b_{-m_k} c_0 c_1 | 0 },
\nonumber \\
  \widetilde{M}_{n_1 m_1 n_2 m_2 \cdots n_k m_k} &\equiv&
  \langle \widetilde{B}'_t | c_{-n_1} b_{-m_1} c_{-n_2} b_{-m_2}
  \cdots c_{-n_k} b_{-m_k} c_0 c_1 | 0 \rangle,
\end{eqnarray}
and
\begin{eqnarray}
  M (z_1, w_1, z_2, w_2, \cdots, z_k, w_k) &\equiv&
  \vev{ B'_t | c(w_1) b(z_1) c(w_2) b(z_2) \cdots
        c(w_k) b(z_k) c_0 c_1 | 0 },
\nonumber \\
  \widetilde{M} (z_1, w_1, z_2, w_2, \cdots, z_k, w_k) &\equiv&
  \langle \widetilde{B}'_t | c'(w_1) b'(z_1) c'(w_2) b'(z_2)
  \cdots c'(w_k) b'(z_k) c_0 | 0' \rangle.
\nonumber \\
\end{eqnarray}
Using (\ref{f^p-identity}), we can show that
\begin{equation}
  M (z_1, w_1, z_2, w_2, \cdots, z_k, w_k)
  = \frac{w_1}{z_1} \frac{w_2}{z_2} \cdots \frac{w_k}{z_k}
  \widetilde{M} (z_1, w_1, z_2, w_2, \cdots, z_k, w_k).
\end{equation}
Therefore, we have
\begin{eqnarray}
  M_{n_1 m_1 n_2 m_2 \cdots n_k m_k}
  &=& \prod_{i=1}^{k} \left[
    \oint \frac{dw_i}{2 \pi i} \frac{1}{w_i^{n_i+2}}
    \oint \frac{dz_i}{2 \pi i} \frac{1}{z_i^{m_i-1}} \right]
    M (z_1, w_1, z_2, w_2, \cdots, z_k, w_k)
\nonumber \\
  &=& \prod_{i=1}^{k} \left[
    \oint \frac{dw_i}{2 \pi i} \frac{1}{w_i^{n_i+1}}
    \oint \frac{dz_i}{2 \pi i} \frac{1}{z_i^{m_i}} \right]
    \widetilde{M} (z_1, w_1, z_2, w_2, \cdots, z_k, w_k)
\nonumber \\
  &=& \widetilde{M}_{n_1 m_1 n_2 m_2 \cdots n_k m_k}.
\end{eqnarray}
This completes the proof that
$\vev{ B'_t | \phi }=\langle \widetilde{B}'_t | \phi \rangle$
for any state $\ket{\phi}$ in the Fock space.

\subsection{Generalization to other surface states}

Our result for the twisted regulated butterfly state can be
generalized to a certain class of other surface states.
The key identity of the proof in the previous subsection
was (\ref{f^p-identity}).
In general, if a surface state $\ket{\Sigma}$ is defined by
a conformal transformation $f (\xi)$ through the relation
\begin{equation}
  \vev{ \phi | \Sigma } = \vev{ f \circ \phi (0) }
\end{equation}
for any state $\ket{ \phi }$ in the Fock space,
where the correlation function is evaluated on an upper-half plane,
and the conformal transformation $f (\xi)$ satisfies
\begin{equation}
  \frac{d f (\xi)}{d \xi} = \frac{1}{C} \frac{f (\xi) - f(0) }{\xi}
  \prod_{i} \left( f(\xi)-p_i \right)^{\alpha_i},
\label{condition-f}
\end{equation}
where $C$ is a factor independent of $\xi$, and
\begin{equation}
  \sum_{i} \alpha_i = 1,
\end{equation}
the twisted surface state $\ket{\Sigma'}$ can be represented
by a correlation function in the ordinary ghost CFT
with $c$-ghost insertions at $p_i$ when $\alpha_i=1$
and $b$-ghost insertions at $p_i$ when $\alpha_i=-1$.
The insertion points do not have to be on the boundary.
Furthermore, we can handle the case where the ghost numbers
$\alpha_i$ take values other than $1$ or $-1$, for example,
by bosonization.

The condition (\ref{condition-f}) can be brought to
a more convenient form
in terms of the inverse function $\xi = f^{-1} (z)$.
It is given by
\begin{equation}
  \frac{d \ln f^{-1} (z)}{dz} = \frac{C}{z-f(0)}
  \prod_{i} (z-p_i)^{-\alpha_i},
\label{condition-f^-1}
\end{equation}
where $C$ is a factor independent of $z$.

Let us first verify the condition (\ref{condition-f^-1})
for the regulated butterfly state.
The inverse function $h_t^{-1} (z)$ is given by
\begin{equation}
  h_t^{-1} (z) = \frac{pz}{\sqrt{(1-(1+pt)z)(1-(1-pt)z)}},
\end{equation}
and the derivative of $\ln h_t^{-1} (z)$ is
\begin{equation}
  \frac{d \ln h_t^{-1} (z)}{dz} = \frac{1}{p^2 t^2 -1}
  \frac{z-1}{z \left( z-\frac{1}{1+pt} \right)
             \left( z-\frac{1}{1-pt} \right)},
\end{equation}
which takes the form of (\ref{condition-f^-1}).

Now consider the wedge state
\cite{Rastelli:2000iu, Kishimoto:2001ac, Schnabl:2002gg}.
It is labeled by a real number $n$
with $n \ge 1$ and defined by the conformal transformation
\begin{equation}
  f_n (\xi) = \frac{n}{2}
  \tan \left( \frac{2}{n} \arctan \xi \right).
\label{wedge-definition}
\end{equation}
In the large $n$ limit, the wedge state becomes a star-algebra
projector, which is called the sliver state.

When $n$ is a positive integer, the wedge state is given by
a star product of $n-1$ vacuum states.
For example, the wedge state with $n=2$ is the vacuum state
$\ket{0}$ itself, and $\ket{0} \ast \ket{0}$ for $n=3$,
$\ket{0} \ast \ket{0} \ast \ket{0}$ for $n=4$, and so on.
In this case, we expect that the twisted wedge state
can be represented in terms of the ordinary ghost CFT
by inserting a $c$ ghost at the midpoint of the boundary
of each vacuum state and by putting an appropriate operator
with a negative ghost number at the string midpoint
which has an excess angle.

Let us examine if the wedge state satisfies
the condition (\ref{condition-f^-1}).
The inverse function $f_n^{-1} (z)$ is given by
\begin{equation}
  f_n^{-1} (z) =
  \tan \left( \frac{n}{2} \arctan \frac{2 z}{n} \right),
\end{equation}
and the derivative of $\ln f_n^{-1} (z)$ is
\begin{equation}
  \frac{d \ln f_n^{-1} (z)}{dz} = \frac{2 n^2}{4 z^2 + n^2}
  \left[ \sin \left( n \arctan \frac{2 z}{n}
  \right) \right]^{-1}.
\label{wedge-1}
\end{equation}
It is not obvious if this can be transformed to the form
(\ref{condition-f^-1}), but when $n$ is an odd positive integer,
it turns out to be the case:
\begin{equation}
  \frac{d \ln f_n^{-1} (z)}{dz}
  = \frac{ n^2 (-1)^{\frac{n-1}{2}}}{2 z}
  \left( z - \frac{ni}{2} \right)^{\frac{n}{2}-1}
  \left( z + \frac{ni}{2} \right)^{\frac{n}{2}-1}
  \prod_{m=1}^{n-1}
  \left( z - \frac{n}{2} \tan \frac{m \pi}{n} \right)^{-1}.
\label{wedge-2}
\end{equation}
A derivation of this expression is given in Appendix B.
As we guessed, if we insert $n-1$ $c$ ghosts at
\begin{equation}
  \frac{n}{2} \tan \frac{m \pi}{n}, \quad
  m = 1, 2, \cdots, n-1,
\end{equation}
which are the midpoints of the boundary of the $n-1$ vacuum states,
and operators with ghost number $1-n/2$ at $\pm ni/2$,
the twisted wedge state can be described by a correlation
function of a CFT with a vanishing central charge
when $n$ is an odd positive integer.
It will be straightforward to generalize the derivation
to the case where $n$ is an even positive integer
by making an appropriate $SL(2,R)$ transformation
to avoid the operator insertion at the infinity.
However, the generalization to the case
where $n$ is not an integer is nontrivial,
and it is not clear if such a description
in terms of a CFT with a vanishing central charge exists.

The operator insertion at the open-string midpoint, 
which corresponds to $\pm ni/2$ under the doubling trick,
implies that the singularity of the twisted sliver state
is not completely resolved in our description
by regularizing it to the twisted wedge state.
In particular, the star multiplication of two twisted wedge states
is not well-defined because of coincident operators
at the open-string midpoint.
The remaining singularity has to be regularized further,
for example, by displacing the operators to $\pm i(n/2+\epsilon)$
with $\epsilon >0$ in our description.\footnote
{
There is a possibility that the singularity from the coincident
operators is canceled by singular conformal factors so that
we obtain a finite result in the limit $\epsilon \to 0$.
We do not know whether the singularity
resulting from the operator insertion
at the open-string midpoint is really harmful
or just an artifact of our description.
However, regularization is necessary as long as
we use our description of the twisted wedge state.
}

Once the remaining singularity is regularized, we can in principle
carry out computations involving the twisted wedge state
in a CFT with a vanishing central charge. However, such computations
would be much more awkward to handle
compared to the case of the twisted regulated butterfly state.
First, when $n$ is an odd integer, the ghost charge of the operators
at $\pm ni/2$, which is $1-n/2$, is fractional
so that we have to bosonize the ghosts.
When $n$ is an even integer, the charge is an integer but a large
negative number if $n$ is large. Therefore, it would be in practice
difficult to handle without bosonization.
Second, the number of operator insertions increases as $n$ becomes
large, and it diverges in the large $n$ limit.
In the case of the twisted regulated butterfly state,
the number of operator insertions is three
for an arbitrary $t$. Furthermore, the open-string
midpoint is as regular as that of a state in the Fock space.
These are the reasons
why we study the twisted regulated butterfly state
among other star-algebra projectors.

\section{Exact computations of various quantities}
\setcounter{equation}{0}

Now we have the representation (\ref{B'_t})
of the twisted regulated butterfly state
$\ket{ B'_t }$ in terms of a CFT with a vanishing central charge,
we can compute various quantities involving this state
without evaluating conformal anomaly.
We are particularly interested in the quantities
$\vev{ B'_t| Q | \phi }$,
$\vev{ B'_t \ast B'_t | \phi }$,
$\vev{ B'_t | Q | B'_t }$,
and  $\vev{ B'_t \ast B'_t | B'_t }$
in the context of vacuum string field theory, where
\begin{equation}
  Q = \frac{1}{2i} (c(i) - c(-i)),
\label{Q}
\end{equation}
and $\ket{\phi}$ is a state in the Fock space.
We will also denote a state in the Fock space by $\ket{\phi}$
throughout the rest of the paper.
Let us compute these quantities in this section.

\subsection{$\vev{ B'_t | Q | \phi }$}

Since the open-string midpoint
of the regulated butterfly state $\ket{B_t}$
is as regular as that of an ordinary state in the Fock space,
the quantities $\vev{ B'_t | Q | \phi }$
and $\vev{ B'_t | Q | B'_t }$ are well-defined for $0 \le t <1$.
We do not need to regularize the midpoint $c$-ghost insertion $Q$
unlike the case of the wedge state.

The operator $Q$ is mapped to
\begin{eqnarray}
  && \frac{i}{2p} \sqrt{1-t^2} \left[
  (1-ip \sqrt{1-t^2})^2 \,
  c \left( \frac{1}{1-ip \sqrt{1-t^2}} \right) \right.
\nonumber \\
  && \qquad \qquad \qquad - (1+ip \sqrt{1-t^2})^2 \, \left.
  c \left( \frac{1}{1+ip \sqrt{1-t^2}} \right) \right]
\end{eqnarray}
by the conformal transformation $h_t (\xi)$ so that
$\vev{ B'_t | Q | \phi }$ is given by
\begin{eqnarray}
  && \vev{ B'_t | Q | \phi }
  = -\frac{it}{4p} (p^2 t^2 -1)^2 \sqrt{1-t^2}
\nonumber \\ &&
  \times \left[ (1-ip \sqrt{1-t^2})^2
  \vev{ c \left( \frac{1}{1+pt} \right) b (1) \,
        c \left( \frac{1}{1-pt} \right) 
        c \left( \frac{1}{1-ip \sqrt{1-t^2}} \right)
        h_t \circ \phi (0) } \right.
\nonumber \\ && \quad
  \left. - (1+ip \sqrt{1-t^2})^2
  \vev{ c \left( \frac{1}{1+pt} \right) b (1) \,
        c \left( \frac{1}{1-pt} \right) 
        c \left( \frac{1}{1+ip \sqrt{1-t^2}} \right)
        h_t \circ \phi (0) } \right].
\nonumber \\
\end{eqnarray}
As a check, it can be easily verified from this expression
that $\vev{ B'_t | Q c_1 | 0 }=1$
for any $t$ in the range $0 \le t <1$.

In the butterfly limit $t \to 1$,
the $c$ ghost coming from $Q$ approaches $b$ so that
the two operators can be replaced by the leading term of the OPE:
\begin{equation}
  b(1) \, c \left( \frac{1}{1 \pm ip \sqrt{1-t^2}} \right)
  = \pm \frac{1}{ip \sqrt{2(1-t)}} + O(1).
\label{Q-B'_t-OPE}
\end{equation}
Therefore, the leading term of $\vev{ B'_t | Q | \phi }$
in the limit $t \to 1$ is finite and given by
\begin{equation}
  \vev{ B'_t | Q | \phi }
  = -\frac{(p^2-1)^2}{2 p^2}
  \vev{ c \left( \frac{1}{1+p} \right)
        c \left( \frac{1}{1-p} \right) 
        h_B \circ \phi (0) } + O(1-t).
\end{equation}
Note that terms of $O(\sqrt{1-t})$ cancel so that
the next-to-leading order is $O(1-t)$.
Note also that
\begin{equation}
  h_t (\xi) = h_B (\xi) + O(1-t).
\end{equation}

\subsection{$\vev{ B'_t \ast B'_t | \phi }$}

Let us denote the conformal transformation associated with
the surface state $\ket{B_t \ast B_t}$ by $\widetilde{f}_t (\xi)$:
\begin{equation}
  \vev{ \phi | B_t \ast B_t }
  = \vev{ \widetilde{f}_t \circ \phi (0) }.
\end{equation}
We also require that $\widetilde{f}_t (0) = 0$
and $\widetilde{f}_t (1) = -\widetilde{f}_t (-1)$
to fix the ambiguity coming from $SL(2,R)$ transformations.
The overall normalization of $\widetilde{f}_t (\xi)$ is still
undermined, but it will also be fixed shortly.

When we deal with the star multiplication
of the regulated butterfly state,
it is convenient to use the $\hat{z}$ coordinate
defined by
\begin{equation}
  \hat{z} = \arctan \xi.
\end{equation}
It was shown in \cite{Gaiotto:2002kf}
that $\hat{z} = \arctan \xi$
and $z = \widetilde{f}_t (\xi)$ are related by
\begin{equation}
  \frac{d \hat{z}}{dz} = \frac{1}{1+ d^2 z^2}
  \frac{1-\beta^2 z^2}
  {\sqrt{(1-\alpha^2 z^2)(1-\gamma^2 z^2)}},
\end{equation}
where $\alpha$, $\beta$, $\gamma$, and $d$ are
functions of $t$.\footnote
{
In \cite{Gaiotto:2002kf}, $\alpha$, $\beta$ and $\gamma$ are
denoted by $a$, $b$, and $c$, respectively.
We have changed the names to avoid possible confusion
with $bc$ ghosts.
In fact, the points $\pm 1/b$ and $\pm 1/c$ are
the positions where $b$ and $c$ ghosts, respectively,
will be inserted.
}
We use this relation
to fix the overall normalization of $\widetilde{f}_t (\xi)$,
namely,
\begin{equation}
  \left. \frac{d \hat{z}}{dz} \right|_{z=0} = 1.
\label{tilde-f-normalization}
\end{equation}
Schnabl derived an explicit form of $\widetilde{f}_t (\xi)$
in \cite{Schnabl:2002ff}.
By appropriately rescaling the expression in \cite{Schnabl:2002ff}
to satisfy (\ref{tilde-f-normalization}),
$\widetilde{f}_t (\xi)$ is given by
\begin{equation}
  \widetilde{f}_t (\xi)
  = \sqrt{ \frac{3}{4} \frac{9-a^2}{(1-a^2)(a^2+3)}
    \left[ \tan^2 \left( \frac{2}{3}
    \arctan \sqrt{\frac{\xi^2 + t^2}{1 + t^2 \xi^2}} \right)
    - \frac{a^2}{3} \right]},
\label{f-tilde}
\end{equation}
where
\begin{equation}
  a \equiv \sqrt{3} \tan \left( \frac{2}{3} \arctan t \right).
\end{equation}
It can be easily verified that $\widetilde{f}_t (\xi)$ reduces to
$f_B (\xi)$ in the limit $t \to 1$:
\begin{equation}
  \widetilde{f}_t (\xi) = f_B (\xi) + O(1-t).
\end{equation}
This shows that the butterfly state is a star-algebra projector.
The functions $\alpha$, $\beta$, $\gamma$, and $d$ were not
determined explicitly in \cite{Gaiotto:2002kf},
but they can be determined from
(\ref{f-tilde}) and given by
\begin{eqnarray}
  && \alpha = \frac{2(1+a)}{3+a} \sqrt{\frac{1+a}{3-a}}, \quad
  \beta = \frac{2}{9-a^2} \sqrt{(1-a^2)(a^2+3)},
\nonumber \\
  && \gamma = \frac{2(1-a)}{3-a} \sqrt{\frac{1-a}{3+a}}, \quad
  d = 2 \sqrt{\frac{1-a^2}{9-a^2}}.
\label{alpha-beta-gamma-d}
\end{eqnarray}

In order to compute $\vev{ B'_t \ast B'_t | \phi }$,
we need to know where the inserted operators for each of
$\ket{B'_t}$ are mapped to
and the conformal factors associated with the mapping.
Let us prepare two $\vev{ B'_t | \phi }$'s in the coordinates
$z_i = f_t (\xi_i)$ where $i=1,2$ to construct
$\vev{ B'_t \ast B'_t | \phi }$ in the coordinate
$z = \widetilde{f}_t (\xi)$.
The relation between $z_i$ and $z$ can be derived
in the $\hat{z}$ representation.
It was shown in \cite{Gaiotto:2002kf} that
$\hat{z}_i = \arctan \xi_i$ and $\hat{z} = \arctan \xi$
coincide up to a possible translation by the amount of $\pm \pi$
in the outside region of the image of the local coordinate.
Therefore, $\tan \hat{z}_i = \tan \hat{z}$ for $i=1,2$.
Since
\begin{equation}
  \tan^2 \hat{z}_i = \frac{z_i^2}{1- t^2 z_i^2}, \quad
  \tan^2 \hat{z}
  = \frac{(1+d^2 z^2)^{\frac{3}{2}}-(1-\frac{4 \beta^2}{d^2} z^2)}
         {(1+d^2 z^2)^{\frac{3}{2}}+(1-\frac{4 \beta^2}{d^2} z^2)},
\label{z-hat-z}
\end{equation}
$z_i$ and $z$ are related by
\begin{equation}
  z_i^2 
  = \frac{(1+d^2 z^2)^{\frac{3}{2}}-(1-\frac{4 \beta^2}{d^2} z^2)}
         {(1+t^2)(1+d^2 z^2)^{\frac{3}{2}}
          +(1-t^2)(1-\frac{4 \beta^2}{d^2} z^2)}
\end{equation}
for both $i=1$ and $i=2$.
The $b$ ghost is inserted at the infinity in the $z_i$ coordinate
so that it is convenient to make an inversion.
Let us introduce the inverted coordinates
$z_i' = I \circ f_t (\xi_i) = -1/z_i$ for $i=1,2$.
They are related to the $z$ coordinate of
$\vev{ B'_t \ast B'_t | \phi }$ by
\begin{equation}
  z'^2 
  = \frac{(1+t^2)(1+d^2 z^2)^{\frac{3}{2}}
          +(1-t^2)(1-\frac{4 \beta^2}{d^2} z^2)}
    {(1+d^2 z^2)^{\frac{3}{2}}-(1-\frac{4 \beta^2}{d^2} z^2)},
\label{z-z'}
\end{equation}
where $z'=z_1'$ or $z'=z_2'$.
As we have presented in (\ref{B'_t-another}),
two $c$ ghosts are inserted at $t$ and $-t$, and one $b$ ghost
is inserted at the origin in the $z'$ coordinate.
It was shown in \cite{Gaiotto:2002kf}
that these points are mapped to
\begin{equation}
  z'=-t \to z = \frac{1}{\alpha}, \quad
  z'= 0 \to z = \frac{1}{\beta}, \quad
  z'= t \to z = \frac{1}{\gamma}, 
\end{equation}
or to
\begin{equation}
  z'=-t \to z = -\frac{1}{\gamma}, \quad
  z'= 0 \to z = -\frac{1}{\beta}, \quad
  z'= t \to z = -\frac{1}{\alpha}, 
\end{equation}
depending on whether $z'=z'_1$ or $z'=z'_2$.
It can be verified that the relation (\ref{z-z'}) is satisfied
when $(z'^2, z^2) = (t^2, 1/\alpha^2)$, $(z'^2,z^2) = (0, 1/\beta^2)$,
and $(z'^2, z^2) = (t^2, 1/\gamma^2)$.

Let us next compute $dz'/dz$ at these points to determine
conformal factors. From (\ref{z-z'}) and the fact that
$dz'/dz > 0$ at these points, it is not
too difficult to derive the following results:
\begin{eqnarray}
  && \left. \frac{dz'}{dz} \right|_{z=\pm \frac{1}{\alpha}}
  = \frac{a(3+a)(1+a)^2}{8t \sqrt{(1+a)(3-a)}}, \quad
  \left. \frac{dz'}{dz} \right|_{z=\pm \frac{1}{\beta}}
  = \frac{\sqrt{(1-t^4)(1-a^2)(3+a^2)}}{4 \sqrt{3}},
\nonumber \\
  && \left. \frac{dz'}{dz} \right|_{z=\pm \frac{1}{\gamma}}
  = \frac{a(3-a)(1-a)^2}{8t \sqrt{(1-a)(3+a)}}.
\end{eqnarray}
Therefore, the inserted operators are mapped from
the $z'$ coordinate to the $z$ coordinate
including their conformal factors as follows:
\begin{eqnarray}
  && c(\mp t) \to \frac{a(3+a)(1+a)^2}{8t \sqrt{(1+a)(3-a)}} \,
                  c \left( \pm \frac{1}{\alpha} \right), \quad
                  \frac{a(3-a)(1-a)^2}{8t \sqrt{(1-a)(3+a)}} \,
                  c \left( \mp \frac{1}{\gamma} \right), \quad
\nonumber \\
  && b(0) \to \frac{48}{(1-t^4)(1-a^2)(3+a^2)} \,
              b \left( \pm \frac{1}{\beta} \right).
\end{eqnarray}

In order to discuss the limit $t \to 1$ and to compare
$\vev{ B'_t \ast B'_t | \phi }$ with $\vev{ B'_t | Q | \phi }$,
it is convenient to make the $SL(2,R)$ transformation $z/(z+p)$.
The operators in the $z$ coordinate are mapped to
\begin{eqnarray}
  && c \left( \pm \frac{1}{\alpha} \right) \to
  \frac{(p \alpha \pm 1)^2}{p \alpha^2} \,
  c \left( \frac{1}{1 \pm p \alpha} \right), \quad
  b \left( \pm \frac{1}{\beta} \right) \to
  \frac{p^2 \beta^4}{(1 \pm p \beta)^4} \,
  b \left( \frac{1}{1 \pm p \beta} \right),
\nonumber \\
  && c \left( \pm \frac{1}{\gamma} \right) \to
  \frac{(p \gamma \pm 1)^2}{p \gamma^2} \,
  c \left( \frac{1}{1 \pm p \gamma} \right).
\end{eqnarray}
Collecting all the conformal factors and taking into account
the normalization factor $-t/2$ in (\ref{B'_t-another}),
$\vev{ B'_t \ast B'_t | \phi }$ is given by
\begin{eqnarray}
  \vev{ B'_t \ast B'_t | \phi }
  &=& \frac{9}{64} \frac{a^4 (1-a^2)(9-a^2)}{t^2 (1-t^4)^2 (3+a^2)^2}
  \frac{\beta^8 (p^2 \alpha^2-1)^2 (1-p^2 \gamma^2)^2}
       {\alpha^4 \gamma^4 (1-p^2 \beta^2)^4}
\nonumber \\ && \times \left\langle
       c \left( \frac{1}{1 + p \alpha} \right)
       b \left( \frac{1}{1 + p \beta} \right)
       c \left( \frac{1}{1 + p \gamma} \right) \right.
\nonumber \\ && \qquad \times \left.
       c \left( \frac{1}{1 - p \gamma} \right)
       b \left( \frac{1}{1 - p \beta} \right)
       c \left( \frac{1}{1 - p \alpha} \right)
       \widetilde{h}_t \circ \phi (0) \right\rangle,
\end{eqnarray}
where we have defined $\widetilde{h}_t (\xi)$ by
\begin{equation}
  \widetilde{h}_t (\xi) = \frac{\widetilde{f}_t (\xi)}
                               {\widetilde{f}_t (\xi) + p}.
\end{equation}

When $\ket{\phi}$ is $c_1 \ket{0}$,
$\vev{ B'_t \ast B'_t | c_1 | 0 }$ is given by
\begin{equation}
  \vev{ B'_t \ast B'_t | c_1 | 0 }_{density}
  = -\frac{9}{256} \frac{1}{t^2 (1-t^4)^2}
    \frac{a^2 (3+a^2)^2 \sqrt{(9-a^2)(3+a^2)}}{1-a^2},
\label{B'_t-B'_t-vacuum}
\end{equation}
where the subscript $density$ denotes that
the quantity is divided by the volume factor of space-time.
We will also use this notation in what follows.
The expression (\ref{B'_t-B'_t-vacuum}) reproduces
the familiar result in the limit $t \to 0$
where $\ket{B'_t}$ reduces to $c_1 \ket{0}$:
\begin{equation}
  \lim_{t \to 0} \vev{ B'_t \ast B'_t | c_1 | 0 }_{density}
  = -\left( \frac{3 \sqrt{3}}{4} \right)^3
\end{equation}

In the butterfly limit $t \to 1$, the $t$-dependent quantities
$a$, $\alpha$, $\beta$, $\gamma$, and $\widetilde{h}_t (\xi)$
behave as follows:
\begin{eqnarray}
  && 1-a \simeq O(1-t), \quad \alpha \to 1, \quad 
  \beta \simeq O(\sqrt{1-t}), \quad
  \gamma \simeq O \left( (1-t)^{\frac{3}{2}} \right),
\nonumber \\
  && \widetilde{h}_t (\xi) = h_B (\xi) + O(1-t).
\end{eqnarray}
Therefore, four of the six ghost insertions approach $1$
in the limit so that they can be replaced by the leading term
in the OPE:
\begin{equation}
  b \left( \frac{1}{1 + p \beta} \right)
  c \left( \frac{1}{1 + p \gamma} \right)
  c \left( \frac{1}{1 - p \gamma} \right)
  b \left( \frac{1}{1 - p \beta} \right)
  = \frac{4 \sqrt{2}}{p^2} + O(1-t).
\label{B'_t-B'_t-OPE}
\end{equation}
Note that terms of $O( \sqrt{1-t} )$ cancel so that
the next-to-leading order is $O(1-t)$.
The leading term of $\vev{ B'_t \ast B'_t | \phi }$
in the limit is given by
\begin{eqnarray}
  \vev{ B'_t \ast B'_t | \phi }
  = \frac{27 \sqrt{6}}{1024} \frac{1}{(1-t)^3}
    \frac{(p^2-1)^2}{p^2}
  \vev{ c \left( \frac{1}{1+p} \right)
  c \left( \frac{1}{1-p} \right) h_B \circ \phi (0) }
\nonumber \\
  {}+ O \left( \frac{1}{(1-t)^2} \right).
\end{eqnarray}
Therefore, the leading term of $\vev{ B'_t \ast B'_t | \phi }$
is proportional to that of $\vev{ B'_t | Q | \phi }$,
\begin{equation}
  \vev{ B'_t \ast B'_t | \phi }
  = -\frac{27 \sqrt{6}}{512} \frac{1}{(1-t)^3}
    \vev{ B'_t | Q | \phi } + O \left( \frac{1}{(1-t)^2} \right),
\end{equation}
but the proportionality constant diverges as $1/(1-t)^3$.
This shows that the twisted regulated butterfly state
satisfies the equation of motion of vacuum string field theory
in the singular butterfly limit
when its kinetic operator ${\cal Q}$ is given by
the midpoint $c$-ghost insertion.
However, the singularity in the proportionality constant
implies that the normalization of the solution
does not have a finite limit if the kinetic operator ${\cal Q}$
is given by $Q$ times a finite factor.
We will discuss more about this result
in the context of vacuum string field theory in Section 4
after computing other quantities in the following subsections.

\subsection{$\vev{ B'_t | Q | B'_t }$}

Because $\ket{B'_t}$ satisfies the Siegel gauge condition,
this quantity reduces to $\vev{ B'_t | c_0 | B'_t }$,
which has already been computed by Schnabl \cite{Schnabl:2002ff}
in the operator formalism by evaluating conformal anomaly.
Nevertheless, it is useful to compute it in our CFT approach,
and in fact our method can be applied to the computation of
$\vev{ B'_t \ast B'_t | B'_t }$ in the next subsection.

Our strategy for computing $\vev{ B'_t | Q | B'_t }$ is
the same as in the case of $\vev{ B'_t \ast B'_t | \phi }$
in the previous subsection.
Gluing of two regulated butterfly states can be easily done
in the $\hat{z}$ representation.
The resulting surface can be mapped to an upper-half plane
by an appropriate conformal transformation.
If we denote the coordinate in the upper-half plane by $z$,
the relation between $z$ and $z' = I \circ f_t (\xi)$
for each of the two regulated butterfly states can be derived
through the $\hat{z}$ representation.
We will present details of this process in Appendix C.
The final relation between $z$ and $z'$ is as follows:
\begin{equation}
  \frac{4 (z'^2-t^2)}{\left[ z'^2-(1+t^2) \right]^2}
  = \frac{(1-z^2)^2-4 q^2 z^2}{4 (1+q^2) z^2},
\label{z-z'-2}
\end{equation}
where
\begin{equation}
  q = \frac{2 t}{1-t^2}.
\label{q}
\end{equation}
   From this relation, 
we can compute
how operators are mapped from the $z'$ coordinate
to the $z$ coordinate.
The operators $b(0)$, $c(\pm t)$ in the $z'$ coordinate
are mapped in the following way:
\begin{eqnarray}
  c(-t) &\to& \frac{1}{2 (1+t^2)} \frac{1-t}{1+t} \,
            c \left( -\frac{1+t}{1-t} \right), \quad
            \frac{1}{2 (1+t^2)} \frac{1+t}{1-t} \,
            c \left( \frac{1-t}{1+t} \right),
\nonumber \\
  b(0) &\to& \frac{4}{1-t^4} \, b(-1), \quad
             \frac{4}{1-t^4} \, b(1),
\nonumber \\
  c(t) &\to& \frac{1}{2 (1+t^2)} \frac{1+t}{1-t} \,
           c \left( -\frac{1-t}{1+t} \right), \quad
           \frac{1}{2 (1+t^2)} \frac{1-t}{1+t} \,
            c \left( \frac{1+t}{1-t} \right).
\end{eqnarray}
The kinetic operator $Q$ is expressed in the $z$ coordinate as
\begin{equation}
  Q \to -\frac{1-t^2}{2 (1+t^2)} \left( c(i)+c(-i) \right).
\end{equation}
Collecting all the conformal factors and taking into account
the normalization factor $-t/2$ in (\ref{B'_t-another}),
$\vev{ B'_t | Q | B'_t }$ is given by
\begin{eqnarray}
  && \vev{ B'_t | Q | B'_t }
  = -\frac{t^2}{8 (1-t^2)(1+t^2)^7}
\nonumber \\ && \qquad \qquad \qquad
  \times \left[ \vev{ c \left( -\frac{1+t}{1-t} \right) b(-1) \,
        c \left( -\frac{1-t}{1+t} \right) c(i) \,
        c \left( \frac{1-t}{1+t} \right) b(1) \,
        c \left( \frac{1+t}{1-t} \right)} \right.
\nonumber \\ && \qquad \qquad \qquad
  + \left. \vev{ c \left( -\frac{1+t}{1-t} \right) b(-1) \,
        c \left( -\frac{1-t}{1+t} \right) c(-i) \,
        c \left( \frac{1-t}{1+t} \right) b(1) \,
        c \left( \frac{1+t}{1-t} \right)} \right].
\nonumber \\
\end{eqnarray}
It is straightforward to evaluate the correlation functions
and the result is given by
\begin{equation}
  \vev{ B'_t | Q | B'_t }_{density}
  = \frac{1}{(1-t^4)^3}.
\end{equation}
We have reproduced the result presented
in Appendix C of \cite{Schnabl:2002ff}
when the parameters $s$ and $\tilde{s}$
of \cite{Schnabl:2002ff}
are given by $s=\tilde{s}=-t^2/2$.

\subsection{$\vev{ B'_t \ast B'_t | B'_t }$}

The computation of $\vev{ B'_t \ast B'_t | B'_t }$
is almost parallel to that of $\vev{ B'_t | Q | B'_t }$
in the previous subsection.
Gluing of three twisted regulated butterfly states
can be done in the $\hat{z}$ representation.
The resulting surface can be mapped to an upper-half plane
by an appropriate conformal transformation.
The derivation of the relation between the $z$ coordinate
of the upper-half plane and the coordinate
$z' = I \circ f_t (\xi)$
for each of the three regulated butterfly states
will be given in Appendix C.
The final relation between $z$ and $z'$ is as follows:
\begin{equation}
  \frac{4 (z'^2-t^2)}{\left[ z'^2-(1+t^2) \right]^2}
  = \frac{z^2 (z^2-3)^2-q^2 (1-3 z^2)^2}{(1+q^2)(1-3 z^2)^2},
\label{z-z'-3}
\end{equation}
where $q$ was defined in (\ref{q}).
The operators $b(0)$, $c(\pm t)$ in the $z'$ coordinate
are mapped to the following operators in the $z$ coordinate:
\begin{eqnarray}
  b(0) &\to&
  \frac{64}{9} \frac{1}{1-t^4} \, b(-\sqrt{3}), \quad
  \frac{4}{9} \frac{1}{1-t^4} \, b(0), \quad
  \frac{64}{9} \frac{1}{1-t^4} \, b(\sqrt{3}),
\nonumber \\
  c(-t) &\to&
  \frac{\sqrt{3}}{16} \frac{a (9-a^2)(1-a)}{t (3+a^2)(1+a)} \,
  c \left( -\frac{3+a}{\sqrt{3} (1-a)} \right), \quad
  \frac{\sqrt{3}}{4} \frac{a (9-a^2)}{t (3+a^2)(1-a^2)} \,
  c \left( -\frac{a}{\sqrt{3}} \right),
\nonumber \\
  && \frac{\sqrt{3}}{16} \frac{a (9-a^2)(1+a)}{t (3+a^2)(1-a)} \,
  c \left( \frac{3-a}{\sqrt{3} (1+a)} \right),
\nonumber \\
  c(t) &\to&
  \frac{\sqrt{3}}{16} \frac{a (9-a^2)(1+a)}{t (3+a^2)(1-a)} \,
  c \left( -\frac{3-a}{\sqrt{3} (1+a)} \right), \quad
  \frac{\sqrt{3}}{4} \frac{a (9-a^2)}{t (3+a^2)(1-a^2)} \,
  c \left( \frac{a}{\sqrt{3}} \right),
\nonumber \\
  && \frac{\sqrt{3}}{16} \frac{a (9-a^2)(1-a)}{t (3+a^2)(1+a)} \,
  c \left( \frac{3+a}{\sqrt{3} (1-a)} \right).
\end{eqnarray}
Collecting all the conformal factors and taking into account
the normalization factor $-t/2$ in (\ref{B'_t-another}),
$\vev{ B'_t \ast B'_t | B'_t }$ is given by
\begin{eqnarray}
  && \vev{ B'_t \ast B'_t | B'_t }
  = -\frac{a^6 (9-a^2)^6}
          {2^9 3^3 t^3 (1-t^4)^3 (3+a^2)^6 (1-a^2)^2}
\nonumber \\
  && \qquad \qquad \qquad \quad \times \left\langle
  c \left( -\frac{3+a}{\sqrt{3} (1-a)} \right) b(-\sqrt{3}) \,
  c \left( -\frac{3-a}{\sqrt{3} (1+a)} \right) \right.
\nonumber \\
  && \qquad \qquad \qquad \quad \times \left.
  c \left( -\frac{a}{\sqrt{3}} \right) b(0) \,
  c \left( \frac{a}{\sqrt{3}} \right)
  c \left( \frac{3-a}{\sqrt{3} (1+a)} \right) b(\sqrt{3}) \,
  c \left( \frac{3+a}{\sqrt{3} (1-a)} \right) \right\rangle.
\nonumber \\
\end{eqnarray}
After calculating the correlation function,
the density of $\vev{ B'_t \ast B'_t | B'_t }$ is given by
\begin{equation}
  \vev{ B'_t \ast B'_t | B'_t }_{density}
  = - \left[ \frac{3 \sqrt{3}}{4}
      \frac{1}{(1-t^2)(1-t^4)} \right]^3.
\end{equation}
The result reproduces the familiar value $-(3 \sqrt{3}/4)^3$
in the limit $t \to 0$ where $\ket{B'_t}$ reduces to $c_1 \ket{0}$.
On the other hand, $\vev{ B'_t \ast B'_t | B'_t }$ diverges
in the butterfly limit $t \to 1$ as $1/(1-t)^6$.

\section{Relation to vacuum string field theory}
\setcounter{equation}{0}

We have provided the representation (\ref{B'_t})
of the twisted regulated butterfly state
in terms of a CFT with a vanishing central charge in Section 2,
and have presented some exact computations of various quantities
relevant to vacuum string field theory in Section 3.
We are now ready to discuss
the issues we raised in the introduction.

\subsection{Vacuum string field theory conjecture}

The action of Witten's cubic open string field theory
\cite{Witten:1985cc} is given by
\begin{equation}
  S = -\frac{1}{\alpha'^3 g_T^2} \left[
      \frac{1}{2} \vev{ \Psi | Q_B | \Psi }
      + \frac{1}{3} \vev{ \Psi | \Psi \ast \Psi } \right],
\end{equation}
where $g_T$ is the on-shell three-tachyon coupling constant.
If we expand the action around the solution $\ket{\Psi_0}$
corresponding to the tachyon vacuum
as $\ket{\Psi}=\ket{\Psi_0}+ | \widetilde{\Psi} \rangle$,
it will take the same form except for the kinetic operator:
\begin{equation}
  S = S_0 -\frac{1}{\alpha'^3 g_T^2} \left[
      \frac{1}{2} \langle \widetilde{\Psi}
      | {\cal Q} | \widetilde{\Psi} \rangle
      + \frac{1}{3} \langle \widetilde{\Psi}
      | \widetilde{\Psi} \ast \widetilde{\Psi} \rangle \right],
\end{equation}
where $S_0$ is the value of the action for $\ket{\Psi_0}$
and ${\cal Q}$ is the kinetic operator at the tachyon vacuum.
It was conjectured in \cite{Rastelli:2000hv}
that ${\cal Q}$ can be made purely
of ghost fields by field redefinition, and string field theory
with this conjectured form of the kinetic operator is called
vacuum string field theory.

A more specific conjecture on ${\cal Q}$ was put forward later
in \cite{Gaiotto:2001ji}.
The kinetic operator ${\cal Q}$ does not seem to be
made purely of ghost fields when we expand the action
around the approximate solution constructed by level truncation.
It was conjectured that there exists a one-parameter family
of field redefinition which takes ${\cal Q}$ to the following
form:
\begin{equation}
  {\cal Q} = \frac{Q}{\epsilon} \left[ 1 + o(\epsilon) \right],
\label{Q-conjecture}
\end{equation}
where $Q$ is the midpoint $c$-ghost insertion
defined in (\ref{Q}), $\epsilon$ is
the parameter of the field redefinition, and
we denoted  $o(\epsilon)$ by terms which vanish in the limit
$\epsilon \to 0$.
In the singular limit $\epsilon \to 0$, the midpoint $c$-ghost
insertion $Q$ dominates in the kinetic operator ${\cal Q}$
with an infinite coefficient.

\subsection{Subleading structure of vacuum string field theory}

If the physics depends on the details of the subleading terms
of the kinetic operator, the vacuum string field theory conjecture
would not be so useful.
We know very little about the subleading terms, and in fact
even the form of the leading term $Q/\epsilon$ is a conjecture.
It is implicitly assumed in the conjecture that
a kind of universality works
as in the case of renormalizable quantum field theory.
If such a universality exists, can we set the subleading
terms to zero and define ${\cal Q}$ as the $\epsilon \to 0$ limit
of $Q/\epsilon$? The action is then given by
\begin{equation}
  S_{\rm leading} = -\frac{1}{\alpha'^3 g_T^2} \left[
      \frac{1}{2 \epsilon} \vev{ \Psi | Q | \Psi}
      + \frac{1}{3} \vev{ \Psi | \Psi \ast \Psi } \right].
\label{S_leading}
\end{equation}

The answer to this question seems to be negative from the following
argument. Unlike the case of Witten's string field theory with
the BRST operator $Q_B$, there exists a field redefinition which
keeps the cubic term intact but changes the normalization of $Q$.
Therefore, if (\ref{S_leading}) is exact and there are no subleading
terms, the value of the coupling constant $g_T$ can be changed
by field redefinition.

Let us demonstrate this explicitly.
It is known that field redefinition generated by
$K_n = L_n - (-1)^n L_{-n}$ preserves the cubic term.
A simple example of field redefinition which changes
the normalization of $Q$ is given by
$\ket{\Psi} = e^{a K_2} | \widetilde{\Psi} \rangle$
since $[K_2, Q] = 4Q$.
Therefore, if we redefine the string field as
$\ket{\Psi} = e^{a (K_2-4)} | \widetilde{\Psi} \rangle$
and write the action in terms of $| \widetilde{\Psi} \rangle$,
the action is multiplied by an overall factor $e^{-12a}$
so that this changes the coupling constant
$g_T$ to $e^{6a} g_T$.
We can construct infinitely many such examples
in terms of a linear combination of $K_n$ because
$[K_{2n}, Q] = -4n (-1)^n Q$.

There is another problem if we assume that
the action (\ref{S_leading}) is exact.
We can absorb not only $g_T$ but also $\epsilon$
by redefinition. For example, if we redefine the string field as
\begin{equation}
  \ket{\Psi} = \alpha' g_T^{\frac{2}{3}}
  \exp \left[ -\left( \frac{1}{4} \ln \epsilon
  + \frac{1}{6} \ln g_T + \frac{1}{4} \ln \alpha'
  \right) K_2 \right] | \widetilde{\Psi} \rangle
\end{equation}
and write the action in terms of $| \widetilde{\Psi} \rangle$,
the action does not contain any parameters.
Since actions with different values of $\epsilon$
are equivalent to each other as long as $\epsilon$ is finite,
the limit $\epsilon \to 0$ does not make sense.
Therefore, we conclude that subleading terms in (\ref{Q-conjecture})
must be present in order for vacuum string field theory to have
a parameter corresponding to the string coupling constant.

\subsection{Finiteness of the energy density}

The twisted regulated butterfly state $\ket{B'_t}$
with an appropriate normalization factor ${\cal N}$ solves
the equation of motion of vacuum string field theory
at the leading order. We are now ready to discuss
the issue of finiteness of the energy density of the solution.

Let us first compute the normalization factor ${\cal N}$.
The equation of motion derived from the leading term of the action
(\ref{S_leading}) consists of two terms.
When $\ket{\Psi} = {\cal N} \ket{B'_t}$, they are given by
\begin{eqnarray}
  \frac{\cal N}{\epsilon} \vev{ B'_t | Q | \phi }
  &=& - \frac{\cal N}{\epsilon} \frac{(p^2-1)^2}{2 p^2}
  \vev{ c \left( \frac{1}{1+p} \right)
        c \left( \frac{1}{1-p} \right) 
        h_B \circ \phi (0) } \left[ 1 + O(1-t) \right],
\nonumber \\
  {\cal N}^2 \vev{ B'_t \ast B'_t | \phi }
  &=& \frac{27 \sqrt{6}}{1024} \frac{{\cal N}^2}{(1-t)^3}
    \frac{(p^2-1)^2}{p^2}
  \vev{ c \left( \frac{1}{1+p} \right)
  c \left( \frac{1}{1-p} \right) h_B \circ \phi (0) }
\nonumber \\
  && \times \left[ 1 + O (1-t) \right].
\end{eqnarray}
Therefore, the equation of motion at the leading order is solved by
\begin{equation}
  {\cal N} = \frac{512}{27 \sqrt{6}} \frac{(1-t)^3}{\epsilon}.
\end{equation}
Note that how we should scale $1-t$ as $\epsilon$ goes to zero
is not determined by the analysis at this order.

Let us next evaluate the energy density. The values of the two terms
in the action (\ref{S_leading}) are given by
\begin{eqnarray}
  -\frac{{\cal N}^2}{2 \alpha'^3 g_T^2 \epsilon}
  \vev{ B'_t | Q | B'_t }_{density}
  &=& -\frac{65536}{2187} \frac{(1-t)^3}
       {\alpha'^3 g_T^2 (1+t)^3 (1+t^2)^3 \epsilon^3}
\nonumber \\
  &\sim& -\frac{1024}{2187} \frac{(1-t)^3}
          {\alpha'^3 g_T^2 \epsilon^3},
\nonumber \\
  -\frac{{\cal N}^3}{3 \alpha'^3 g_T^2}
  \vev{ B'_t \ast B'_t | B'_t }_{density}
  &=& \frac{524288 \sqrt{2}}{2187} \frac{(1-t)^3}
      {\alpha'^3 g_T^2 (1+t)^6 (1+t^2)^3 \epsilon^3}
\nonumber \\
  &\sim& \frac{1024 \sqrt{2}}{2187}
         \frac{(1-t)^3}{\alpha'^3 g_T^2 \epsilon^3},
\label{energy-density}
\end{eqnarray}
where we have also presented their leading behavior when $t \to 1$.
Therefore, both terms can be made finite simultaneously
if $1-t$ scales as $\epsilon$ in the limit $\epsilon \to 0$.
However, it should be noted that we cannot make such a scaling
by hand.
Since we are solving a nonlinear equation, the energy density of
the solution must be determined for a given value of $\epsilon$
if we take into account subleading terms.
The fact that the scaling between $\epsilon$ and $1-t$ is not
determined by the analysis at the leading order will probably
be related to the property of $Q$ that its normalization can be
changed by field redefinition.
Therefore, it is encouraging that our result (\ref{energy-density})
admits a limit which makes both terms finite simultaneously,
but whether the energy density
is finite or not depends on the subleading structure
of vacuum string field theory.

\subsection{Possible relevance of the subleading terms}

Even if we assume that the two terms in (\ref{S_leading})
remain finite in the limit $\epsilon \to 0$,
it is still a nontrivial question whether or not
the equation of motion is satisfied
when it is contracted with the solution itself
in the limit:
\begin{equation}
  \vev{ \Psi | {\cal Q} | \Psi }
  + \vev{ \Psi | \Psi \ast \Psi } = 0.
\end{equation}
To make the point clearer, let us introduce the following quantity:
\begin{equation}
  \frac{\vev{\Psi \ast \Psi | \phi}}{\vev{\Psi | {\cal Q} | \phi}}
  \frac{\vev{\Psi | {\cal Q} | \Psi}}{\vev{\Psi \ast \Psi | \Psi}}.
\end{equation}
If this quantity is different from $1$, the equation of motion
contracted with the solution itself is not compatible with
the one contracted with a state $\ket{\phi}$ in the Fock space.
This quantity is independent of the normalizations of
${\cal Q}$, $\ket{\Psi}$, and $\ket{\phi}$ so that
if ${\cal Q}$ is dominated by $Q/\epsilon$
and $\ket{\Psi}$ is given by ${\cal N} \ket{B'_t}$
in the limit, the quantity reduces to
\begin{equation}
  \lim_{t \to 1}
  \frac{\vev{B'_t \ast B'_t | \phi}}{\vev{B'_t | Q | \phi}}
  \frac{\vev{B'_t | Q | B'_t}}{\vev{B'_t \ast B'_t | B'_t}}.
\label{ratio}
\end{equation}
  From the results in Section 3, it is given by
\begin{equation}
  \lim_{t \to 1}
  \frac{\vev{B'_t \ast B'_t | \phi}}{\vev{B'_t | Q | \phi}}
  \frac{\vev{B'_t | Q | B'_t}}{\vev{B'_t \ast B'_t | B'_t}}
  = \frac{\sqrt{2}}{3}.
\label{discrepancy}
\end{equation}
It is finite and independent of $\ket{\phi}$ in the limit,
but different from $1$.
Therefore, the equations
$\vev{ \Psi | {\cal Q} | \Psi } + \vev{ \Psi | \Psi \ast \Psi } = 0$
and
$\vev{ \phi | {\cal Q} | \Psi } + \vev{ \phi | \Psi \ast \Psi } = 0$
for a state $\ket{\phi}$ in the Fock space are not compatible
if we assume that the kinetic operator is dominated
by the midpoint ghost insertion and that the solution is dominated by
the twisted regulated butterfly state.
This conclusion holds whatever scaling limit
we may take for $\epsilon$, $t$, and ${\cal N}$.

What does this result imply? We do not think that it is an immediate
problem of the vacuum string field theory conjecture.
First of all, we have chosen the butterfly state among
infinitely many star-algebra projectors and regularized it
in a particular way. There might be a better choice of a regulated
star-algebra projector, although it is generally expected that
all the star-algebra projectors are formally equivalent.
Another possibility is that subleading terms in the kinetic
operator ${\cal Q}$ contribute to $\vev{ \Psi | {\cal Q} | \Psi }$
at the same order as the leading term.
Possible subleading terms in the solution $\ket{\Psi}$ may also
contribute to $\vev{ \Psi | {\cal Q} | \Psi }$
or $\vev{ \Psi | \Psi \ast \Psi }$ at the same order.
When we say that ${\cal Q}$ is dominated by $Q/\epsilon$
as in (\ref{Q-conjecture}),
we mean that $\vev{\phi_1 | {\cal Q} | \phi_2}$
is dominated by $\vev{\phi_1 | Q | \phi_2}/\epsilon$
for any arbitrary pair of states $\ket{\phi_1}$ and $\ket{\phi_2}$
in the Fock space.
However, the twisted regulated butterfly state is not
in the Fock space so that $\vev{B'_t | Q | B'_t}/\epsilon$
may not dominate in $\vev{B'_t | {\cal Q} | B'_t}$.
We will demonstrate that this is in fact possible.

Let us introduce the following operator:
\begin{equation}
  Q_{\eta}
  = e^{\frac{\eta}{2} L_0} Q e^{\frac{\eta}{2} L_0}.
\end{equation}
This operator cannot be regarded
as a possible regularization of $Q$
because this does not satisfy the requirements
for a kinetic operator of string field theory.
For example, $Q_{\eta}^2$ does not vanish.
We use this operator only to demonstrate
possible relevance of subleading terms in the quantity
$\vev{B'_t | Q_{\eta} | B'_t}$.
Let us first consider the quantity $\vev{B'_t | Q_{\eta} | \phi}$.
It is more or less obvious that $\vev{B'_t | Q_{\eta} | \phi}$
reduces to $\vev{B'_t | Q | \phi}$ in the limit $t \to 1$,
$\eta \to 0$ with the constraint $e^{\eta/2} t < 1$.\footnote
{
The constraint is necessary to avoid a singularity
which occurs when the $c$ ghost from $Q_{\eta}$
and the $b$ ghost from $\ket{B'_t}$ coincide.
}
We can also confirm this from an explicit expression
of $\vev{B'_t | Q_{\eta} | \phi}$:
\begin{eqnarray}
  \vev{ B'_t | Q_{\eta} | \phi }
  &=& -\frac{it}{4p} \frac{(e^{2 \eta} p^2 t^2 -1)^2
       \sqrt{1-e^{\eta} t^2}}{e^{\frac{\eta}{2}}}
\nonumber \\ &&
  \times \left[ (1-ip e^{\frac{\eta}{2}} \sqrt{1-e^{\eta} t^2})^2
  \left\langle c \left( \frac{1}{1+e^{\eta} pt} \right) b (1) \,
        c \left( \frac{1}{1-e^{\eta} pt} \right) \right. \right.
\nonumber \\ && \qquad \qquad \qquad \qquad \qquad \qquad \times
  \left. c \left( \frac{1}
          {1-ip e^{\frac{\eta}{2}} \sqrt{1-e^{\eta} t^2}} \right)
        h_{e^{\eta} t} \circ \phi (0) \right\rangle
\nonumber \\ && \quad
  - (1+ip e^{\frac{\eta}{2}} \sqrt{1-e^{\eta} t^2})^2
  \left\langle c \left( \frac{1}{1+e^{\eta} pt} \right) b (1) \,
        c \left( \frac{1}{1-e^{\eta} pt} \right) \right.
\nonumber \\ && \qquad \qquad \qquad \qquad \qquad \qquad \times
  \left. \left. c \left( \frac{1}
          {1+ip e^{\frac{\eta}{2}} \sqrt{1-e^{\eta} t^2}} \right)
        h_{e^{\eta} t} \circ \phi (0) \right\rangle \right].
\nonumber \\
\end{eqnarray}
We can also compute $\vev{B'_t | Q_{\eta} | B'_t}$ exactly.
When $e^{\frac{\eta}{2} L_0}$ acts on $\ket{B'_t}$,
it just changes the parameter $t$ to $e^{\eta/2} t$
and the normalization of the state.
It is not difficult to show that
\begin{equation}
  e^{\frac{\eta}{2} L_0} \ket{B'_t}
  = e^{-\frac{\eta}{2}} | B'_{e^{\eta/2} t} \rangle.
\end{equation}
Therefore, $\vev{B'_t | Q_{\eta} | B'_t}$ is given by
\begin{equation}
  \vev{B'_t | Q_{\eta} | B'_t}_{density}
  = e^{-\eta} \langle B'_{e^{\eta/2} t}
  | Q | B'_{e^{\eta/2} t} \rangle_{density}
  = \frac{1}{e^{\eta} (1-e^{2 \eta} t^4)^3}.
\end{equation}
We can see from this expression that subleading terms can contribute
at the same order as the leading term.
Let us look at the next-to-leading order.
By comparing terms of $O(\eta)$ on both sides, we find\footnote
{
This quantity has been computed
in Appendix C of \cite{Schnabl:2002ff}
in the operator formalism.
}
\begin{equation}
  \vev{B'_t | (L_0 Q + Q L_0) | B'_t}_{density}
  = \frac{12 t^4}{(1-t^4)^4} - \frac{2}{(1-t^4)^3}.
\end{equation}
It behaves as $1/(1-t)^4$ when $t \to 1$, which is more singular
than the behavior $1/(1-t)^3$ of $\vev{B'_t | Q | B'_t}$.
Therefore, if $\eta$ scales as $1-t$, the next-to-leading term
$\vev{B'_t | (L_0 Q + Q L_0) | B'_t}$ contribute at the same order
as the leading term $\vev{B'_t | Q | B'_t}$.
Higher-order terms in $\eta$ also contribute at the same order
under this scaling.
If we compute the quantity (\ref{ratio})
with $Q$ replaced by $Q_{\eta}$,
the result can be different from $\sqrt{2}/3$.
When $\eta = 2a(1-t)$, for example, it is given by
\begin{equation}
  \lim_{t \to 1}
  \frac{\vev{B'_t \ast B'_t | \phi}}
       {\langle B'_t | Q_{2a(1-t)} | \phi \rangle}
  \frac{\langle B'_t | Q_{2a(1-t)} | B'_t \rangle}
       {\vev{B'_t \ast B'_t | B'_t}}
  = \frac{\sqrt{2}}{3} \frac{1}{(1-a)^3}.
\end{equation}

As can be seen from this example, the quantity
$\vev{\Psi | {\cal Q} | \Psi}$ may not be saturated by
$\vev{\Psi | Q | \Psi}/\epsilon$, and this might be the origin
of the discrepancy between
the value $\sqrt{2}/3$ in (\ref{discrepancy})
and the expected value $1$.
If this is the case, the factorization of the matter and ghost
sectors at the leading order of vacuum string field theory
might be ruined by the subleading terms of the kinetic operator.

\section{Conclusion and discussion}
\setcounter{equation}{0}

We have presented the description of the twisted regulated butterfly
state $\ket{B'_t}$ in terms of a CFT with a vanishing central charge
given by (\ref{B'_t}).
This description enabled us to carry out exact computations
involving the twisted regulated butterfly state
without evaluating conformal anomaly.
As is emphasized in \cite{Asakawa:1998dv},
the generalized gluing and resmoothing theorem
\cite{LeClair:1988sj}
holds only when the total central charge vanishes.
We can now make use of this theorem for computations
involving the twisted regulated butterfly state
using our description.
Our method can also be applied
to a class of other twisted surface states
which satisfy the condition (\ref{condition-f^-1}).

We have derived exact expressions of the quantities
$\vev{ B'_t | Q | \phi }$, $\vev{ B'_t \ast B'_t | \phi }$,
$\vev{ B'_t | Q | B'_t }$,
and $\vev{ B'_t \ast B'_t | B'_t }$ for any state $\ket{\phi}$
in the Fock space in Section 3.
We have provided some analytic formulas regarding the regulated
butterfly state such as the explicit expressions of the parameters
(\ref{alpha-beta-gamma-d}) introduced in \cite{Gaiotto:2002kf},
and the exact relations between coordinates (\ref{z-z'}),
(\ref{z-z'-2}), and (\ref{z-z'-3}).

In our description of the twisted regulated butterfly state,
the way it solves the equation of motion of vacuum string
field theory at the leading order can be understood
through the OPE's (\ref{Q-B'_t-OPE}) and (\ref{B'_t-B'_t-OPE}).
The twisted regulated butterfly state is represented
as the regulated butterfly state with the three operators
insertions $c$, $b$, and $c$ along the boundary with this ordering.
In the computation of $\vev{ B'_t | Q | \phi }$,
the $c$ ghost from $Q$ approaches the $b$ ghost in $\ket{B'_t}$
and replaces it by the identity operator, which is the leading
term of the OPE (\ref{Q-B'_t-OPE}).
In the computation of $\vev{ B'_t \ast B'_t | \phi }$,
six ghosts $c$, $b$, $c$, $c$, $b$, and $c$ are inserted
along the boundary of the glued surface with this ordering.
Four of them, $b$, $c$, $c$, and $b$ approach the midpoint
of the boundary and are replaced by the identity operator,
which is the leading term of the OPE (\ref{B'_t-B'_t-OPE}).
In both cases, we end up with the same state at the leading order
in the limit $t \to 1$ given by the butterfly state
with two $c$-ghost insertions on the boundary.

Once we understand this mechanism, we can construct different
solutions of the equation of motion of vacuum string field theory
by appropriately inserting operators into the regulated butterfly
state. Furthermore, we can also use the same strategy for solving
the equation of motion of Witten's string field theory, or
even solving the equations of motion of Witten's and vacuum
string field theories simultaneously \cite{Okawa:2003zc}.

With the exact results obtained in Section 3,
we have also discussed the issue of finiteness of the energy
density of the solution of vacuum string field theory.
We first argued that subleading terms in the kinetic operator
${\cal Q}$ are necessary in order to have a parameter
corresponding to the string coupling constant.
We then found that there exists a scaling limit of
the parameters $\epsilon$ of vacuum string field theory
and $t$ of the regulated butterfly state which gives
a finite energy density, but whether or not this scaling limit
is realized depends on the subleading terms of ${\cal Q}$.
Finally, we found that the equation of motion contracted with
the solution itself is not compatible with that contracted with
a state in the Fock space if we assume that the midpoint ghost
insertion $Q/\epsilon$ and the twisted regulated butterfly state
dominate in the quantities $\vev{ \Psi | {\cal Q} | \Psi }$
and $\vev{ \Psi | \Psi \ast \Psi }$.
We demonstrated that it is indeed possible for subleading terms
of ${\cal Q}$ to contribute at the same order as the leading term
in $\vev{ \Psi | {\cal Q} | \Psi }$.
Unfortunately, we know very little about the subleading terms
of ${\cal Q}$. In fact, it is a nontrivial problem to construct
a consistent next-to-leading term to be added to the leading term
given by $Q/\epsilon$.\footnote
{
An interesting family of kinetic operators
of cubic string field theory
were recently constructed \cite{Takahashi:2002ez}
and studied \cite{Kishimoto:2002xi, Drukker:2002ct, Drukker:2003hh,
Takahashi:2003pp, Takahashi:2003xe}.
}
Constructing a solution to the equation of motion of vacuum string
field theory with a finite energy density by taking the singular
limit of the regulated butterfly state seems rather subtle.
Our approach presented in this paper enables us to study
this important issue quantitatively, and we believe that
it will be useful for further investigation in the future.

\section*{Acknowledgements}
I would like to thank Takuya Okuda, Hirosi Ooguri,
Martin Schnabl, and Barton Zwiebach for useful discussions.
This work was supported in part by
the DOE grants DF-FC02-94ER40818 (MIT)
and DE-FG03-92ER40701 (Caltech),
and by a McCone Fellowship in Theoretical Physics from
California Institute of Technology.


\appendix
\renewcommand{\thesection}{Appendix \Alph{section}.}
\renewcommand{\theequation}{\Alph{section}.\arabic{equation}}

\section{Conformal field theory formulation of string field theory}
\setcounter{equation}{0}

In the CFT formulation of string field theory
\cite{LeClair:1988sp, LeClair:1988sj},
an open string field is represented
as a wave functional obtained by a path integral
over a certain region in a Riemann surface.
For example, a state $\ket{\phi}$ in the Fock space
can be represented as a wave functional on the arc
$| \xi |=1$ in an upper-half complex plane of $\xi$
by path-integrating over the interior of
the upper half of the unit disk $| \xi | < 1$
with the corresponding operator $\phi (0)$
inserted at the origin
and with the boundary condition of the open string
imposed on the part of the real axis $-1 \le \xi \le 1$.
A more general class of states
such as the regulated butterfly state
can be defined by a path integral over a different region of
a Riemann surface with a boundary
and with possible operator insertions.
When we parameterize the open string on the arc
as $\xi = e^{i \theta}$ with $0 \le \theta \le \pi$,
we refer to the region $\pi/2 \le \theta \le \pi$
as the left half of the open string,
and to the region $0 \le \theta \le \pi/2$
as the right half of the open string.
We also refer to the point $\theta=\pi/2$
as the open-string midpoint.

We use the standard definitions \cite{Witten:1985cc} of
the inner product $\vev{\phi_1 | \phi_2}$
and the star product $\ket{\phi_1 \ast \phi_2}$.
The state $\ket{\phi_1 \ast \phi_2}$ is defined by
gluing together
the right half of the open string of $\ket{\phi_1}$
and the left half of the open string of $\ket{\phi_2}$.
Gluing can be made by conformal transformations
which map the two regions to be glued together
into the same region.
The inner product $\vev{\phi_1 | \phi_2}$ is defined
by gluing the left and right halves of
the open string of $\ket{\phi_1 \ast \phi_2}$.

We use the doubling trick throughout the paper.
For example, $bc$ ghosts on an upper-half plane
are extended to the lower-half plane by
$c (\bar{z}) = \tilde{c} (z)$ and $b (\bar{z}) = \tilde{b} (z)$.
The normalization of correlation functions of
the $bc$ ghost system is given by
\begin{equation}
  \vev{ c(z_1) \, c(z_2) \, c(z_3) }_{ghost}
  = ( z_1 - z_2 )( z_1 - z_3 )( z_2 - z_3 ).
\end{equation}
In the case of a flat space-time in 26 dimensions,
the matter sector of correlation functions is
normalized as follows:
\begin{equation}
  \vev{ 1 }_{matter} = \int d^{26} x.
\end{equation}
In this paper, we only consider correlation functions
which are independent of space-time coordinates so that
the space-time volume always factors out.
The density of the correlation function of three $c$ ghosts
in the whole system which consists of the matter and ghost sectors
is given by
\begin{equation}
  \vev{ c(z_1) c(z_2) c(z_3) }_{density}
  = ( z_1 - z_2 )( z_1 - z_3 )( z_2 - z_3 ).
\label{three-c}
\end{equation}

The normalization of $\ket{\phi}$ is fixed by the condition
that the $SL(2,R)$-invariant vacuum $\ket{0}$
corresponds to the identity operator. From the normalization
of correlation functions (\ref{three-c})
and the mode expansion (\ref{c_n}),
the normalization of the inner product is then fixed as follows:
\begin{equation}
  \vev{ 0 | c_{-1} c_{0} c_{1} | 0 }_{density} = 1,
\end{equation}
where the inner product has been divided by the space-time volume
as denoted by the subscript $density$.

\section{Derivation of (\ref{wedge-2})}
\setcounter{equation}{0}

We will derive the expression (\ref{wedge-2}) from (\ref{wedge-1}),
\begin{eqnarray}
  && \frac{d \ln f_n^{-1} (z)}{dz} = \frac{2 n^2}{4 z^2 + n^2}
  \left[ \sin \left( n \arctan \frac{2 z}{n}
  \right) \right]^{-1}
\nonumber \\
  && = \frac{ n^2 (-1)^{\frac{n-1}{2}}}{2 z}
  \left( z - \frac{ni}{2} \right)^{\frac{n}{2}-1}
  \left( z + \frac{ni}{2} \right)^{\frac{n}{2}-1}
  \prod_{m=1}^{n-1}
  \left( z - \frac{n}{2} \tan \frac{m \pi}{n} \right)^{-1},
\label{wedge-identity}
\end{eqnarray}
when $n$ is an odd positive integer.
Using the identity
\begin{equation}
  \sin nx = 2^{n-1} \prod_{m=1}^{n}
  \sin \left[ x + \frac{(m-1) \pi}{n} \right],
\end{equation}
we can decompose the sine factor in (\ref{wedge-identity})
as follows:
\begin{equation}
  \sin \left( n \arctan \frac{2 z}{n} \right) 
  = 2^{n-1} \prod_{m=1}^{n}
  \sin \left[ \arctan \frac{2 z}{n} + \frac{(m-1) \pi}{n} \right].
\label{sin-factors}
\end{equation}
Since
\begin{equation}
  \sin \left( \arctan x +a \right)
  = \frac{\cos a}{\sqrt{1+x^2}} \left( x + \tan a \right),
\end{equation}
which follows from
\begin{equation}
  \sin ( \arctan x ) = \frac{x}{\sqrt{1+x^2}},  \quad
  \cos ( \arctan x ) = \frac{1}{\sqrt{1+x^2}},
\end{equation}
each factor in (\ref{sin-factors}) can be rewritten as
\begin{equation}
  \sin \left[ \arctan \frac{2 z}{n} + \frac{(m-1) \pi}{n} \right]
  = \frac{\cos \frac{(m-1) \pi}{n}}
         {\sqrt{z^2 + \frac{n^2}{4}}}
    \left[ z + \frac{n}{2} \tan \frac{(m-1) \pi}{n} \right].
\end{equation}
If we write $n = 2 k + 1$, the product of the cosine factors
can be evaluated as
\begin{eqnarray}
  \prod_{m=1}^{n} \cos \frac{(m-1) \pi}{n}
  &=& \prod_{m=1}^{2 k} \cos \frac{m \pi}{2 k +1}
  = (-1)^{k} \left( \prod_{m=1}^{k}
    \cos \frac{m \pi}{2 k +1} \right)^2
\nonumber \\
  &=& \frac{(-1)^{k}}{2^{2 k}}
  = \frac{(-1)^{\frac{n-1}{2}}}{2^{n-1}},
\end{eqnarray}
where we have used the identity
\begin{equation}
  \prod_{m=1}^{k} \cos \frac{m \pi}{2 k +1}
  = \frac{1}{2^{k}}.
\end{equation}
Since
\begin{eqnarray}
  \prod_{m=1}^n
  \left[ z + \frac{n}{2} \tan \frac{(m-1) \pi}{n} \right]
  = z \prod_{m'=1}^n
  \left( z - \frac{n}{2} \tan \frac{m' \pi}{n} \right)
\end{eqnarray}
with $m'=n-m+1$, we have
\begin{eqnarray}
  && \frac{2 n^2}{4 z^2 + n^2}
  \left[ \sin \left( n \arctan \frac{2 z}{n}
  \right) \right]^{-1}
\nonumber \\
  && = \frac{ n^2 (-1)^{\frac{n-1}{2}}}{2 z}
  \left( z^2 + \frac{n^2}{4} \right)^{\frac{n}{2}-1}
  \prod_{m=1}^{n-1}
  \left( z - \frac{n}{2} \tan \frac{m \pi}{n} \right)^{-1}
\nonumber \\
  && = \frac{ n^2 (-1)^{\frac{n-1}{2}}}{2 z}
  \left( z - \frac{ni}{2} \right)^{\frac{n}{2}-1}
  \left( z + \frac{ni}{2} \right)^{\frac{n}{2}-1}
  \prod_{m=1}^{n-1}
  \left( z - \frac{n}{2} \tan \frac{m \pi}{n} \right)^{-1}.
\end{eqnarray}

\section{Derivation of (\ref{z-z'-2}) and (\ref{z-z'-3})}
\setcounter{equation}{0}

We will derive (\ref{z-z'-2}) and (\ref{z-z'-3})
in this appendix.

When we construct $\vev{ B'_t | Q | B'_t }$ and
$\vev{ B'_t \ast B'_t | B'_t }$, we have to glue
two and three regulated butterfly states, respectively.
Gluing them can be easily done
in the $\hat{z}$ representation as in the case of
$\vev{ B'_t \ast B'_t | \phi }$ in Subsection 3.2.
We then need to derive conformal transformations
which map the glued surfaces in the $\hat{z}$ representation
to an upper-half plane.

Let us first consider a simpler problem.
Prepare a regulated butterfly state in the $\hat{z}$ representation,
and glue the left half of the open string with the right half.
The resulting surface can be represented as
the region $| {\rm Re} \, \hat{z} | \le \pi/4 $
of an upper-half plane
with a boundary running from $-\pi/4+i \, {\rm arctanh} \, t$ to
$\pi/4+i \, {\rm arctanh} \, t$ as
\begin{equation}
  -\frac{\pi}{4}+i \, {\rm arctanh} \, t \to -\frac{\pi}{4}
  \to \frac{\pi}{4} \to \frac{\pi}{4}+i \, {\rm arctanh} \, t,
\end{equation}
and with $-\pi/4+ iy$ being identified with
$\pi/4+ iy$ for $y \ge {\rm arctanh} \, t$.
A conformal transformation which maps this surface
to an upper-half plane can be easily found because
the structure of this surface is the same as a surface
for $\vev{\phi | B'_t}$ with a state $\ket{\phi}$
in the Fock space. Therefore, it takes the following form:
\begin{equation}
  \hat{z} = A \arctan
  \left( \frac{B z}{\sqrt{1- q^2 z^2}} \right), \quad
  \frac{d \hat{z}}{dz}
  = \frac{AB}{\left( 1+(B^2-q^2)z^2 \right) \sqrt{1- q^2 z^2}},
\end{equation}
where $A$, $B$, and $q$ are parameters to be determined.
We choose $z= i$ to be the open-string midpoint which
corresponds to the infinity in the $\hat{z}$ coordinate.
Then it follows from the argument in \cite{Gaiotto:2002kf} that
$d \hat{z}/dz$ must have a pole at $z=\pm i$.
This condition determines that $B=\sqrt{1+q^2}$.
The parameter $A$ is fixed by the condition
\begin{eqnarray}
  \frac{\pi}{2} = \int_{-\infty}^{\infty} dz \,
  \frac{d \hat{z}}{dz}
  = \int_{-\infty}^{\infty} dz \,
  \frac{A \sqrt{1+q^2}}{(1+z^2) \sqrt{1- q^2 z^2}}.
\end{eqnarray}
By evaluating the residue at the pole $z=i$
in $d \hat{z}/dz$, this condition determines that $A=1/2$.
The parameter $q$ is determined in terms of $t$ by the condition
that the point $\hat{z} = \pi/4 + i \, {\rm arctanh} \, t$
should be mapped to the infinity in the $z$ coordinate. From
\begin{equation}
  z^2 = \frac{\tan^2 2 \hat{z}}{1+q^2+q^2 \tan^2 2 \hat{z}},
\end{equation}
it follows that
\begin{equation}
  1+q^2+q^2 \tan^2 \left( \frac{\pi}{2}
  + 2i \, {\rm arctanh} \, t \right) = 0.
\end{equation}
Therefore, $q$ is given by $2t/(1-t^2)$.
To summarize, the conformal transformation is determined to be
\begin{equation}
  \hat{z} = \frac{1}{2}
  \arctan \frac{\sqrt{1+q^2} z}{\sqrt{1-q^2 z^2}}
\label{simpler}
\end{equation}
with
\begin{equation}
  q = \frac{2t}{1-t^2}.
\end{equation}

The conformal transformations associated with
$\vev{ B'_t | Q | B'_t }$ and $\vev{ B'_t \ast B'_t | B'_t }$
can be constructed from (\ref{simpler}) by the following trick.
In the case of $\vev{ B'_t | Q | B'_t }$,
let us consider the conformal transformation
associated with a wedge state
(\ref{wedge-definition}) with $n=4$:
\begin{equation}
  z = \tan \left( \frac{1}{2} \arctan \tilde{z} \right),
\end{equation}
where a different normalization has been chosen
such that $\tilde{z}=i$ is mapped to $z=i$.
If the coordinate $\tilde{z}$ is mapped to
$\hat{z}$ by the map (\ref{simpler}), the coordinate $z$ is mapped
to the surface associated with $\vev{ B'_t | Q | B'_t }$.
Since
\begin{equation}
  \tilde{z} = \tan ( 2 \arctan z)
  = \frac{2z}{1-z^2} \equiv g_2 (z),
\end{equation}
the map from $z$ to $\hat{z}$ is given by
\begin{equation}
  \hat{z} = \frac{1}{2}
  \arctan \frac{\sqrt{1+q^2} g_2 (z)}{\sqrt{1-q^2 g_2 (z)^2}}
  = \frac{1}{2} \arctan \left( \frac{2 \sqrt{1+q^2} z}
    {\sqrt{(1-z^2)^2-4 q^2 z^2}} \right).
\label{z-z-hat-2}
\end{equation}

In the case of $\vev{ B'_t \ast B'_t | B'_t }$,
consider the conformal transformation
associated with a wedge state with $n=6$ and make an inversion:
\begin{equation}
  z = -\cot \left( \frac{1}{3} \arctan \tilde{z} \right),
\end{equation}
where the normalization has been again chosen such that $\tilde{z}=i$
is mapped to $z=i$. The purpose of the inversion is to avoid
an operator insertion at $z=\infty$.
If the coordinate $\tilde{z}$ is mapped to
$\hat{z}$ by the map (\ref{simpler}), the coordinate $z$ is mapped
to the surface associated with $\vev{ B'_t \ast B'_t | B'_t }$.
Since
\begin{equation}
  \tilde{z} = -\tan ( 3 \, {\rm arccot} \, z)
  = \frac{1-3z^2}{z^3-3z} \equiv g_3 (z),
\end{equation}
the map from $z$ to $\hat{z}$ is given by
\begin{equation}
  \hat{z} = \frac{1}{2}
  \arctan \frac{\sqrt{1+q^2} g_3 (z)}{\sqrt{1-q^2 g_3 (z)^2}}
  = \frac{1}{2} \arctan \left( \frac{\sqrt{1+q^2} (1-3z^2)}
    {\sqrt{z^2 (z^2-3)^2-q^2 (1-3z^2)^2}} \right).
\label{z-z-hat-3}
\end{equation}

Let us next consider the relation between the $\hat{z}$ coordinate
for either $\vev{ B'_t | Q | B'_t }$
or $\vev{ B'_t \ast B'_t | B'_t }$
and the $\hat{z}_i$ coordinate for each of $\ket{B'_t}$,
namely, $i=1,2$ for $\vev{ B'_t | Q | B'_t }$ and $i=1,2,3$ for
$\vev{ B'_t \ast B'_t | B'_t }$. They are related by
$\hat{z} = \hat{z}_i + m \pi/4$ with an appropriate odd integer $m$
so that
\begin{equation}
  \tan^2 2 \hat{z}_i = \cot^2 2 \hat{z}. 
\label{z-hat-z-hat-appendix}
\end{equation}
   From the relation between $\tan^2 \hat{z}_i$
and the $z_i$ coordinate in (\ref{z-hat-z}),
$\tan^2 2 \hat{z}_i$ is given by
\begin{equation}
  \tan^2 2 \hat{z}_i
  = \left( \frac{2 \tan \hat{z}_i}{1-\tan^2 \hat{z}_i} \right)^2
  = \frac{4 z_i^2 (1-t^2 z_i^2)}{[1-(1+t^2) z_i^2]^2}
  = \frac{4 (z'^2-t^2)}{[z'^2-(1+t^2)]^2},
\label{z-hat-z'}
\end{equation}
where we have introduced $z'_i=-1/z_i$ as in Subsection 3.2
and $z'$ stands for one of the $z'_i$'s.
  From (\ref{z-z-hat-2}), (\ref{z-z-hat-3}),
(\ref{z-hat-z-hat-appendix}), and (\ref{z-hat-z'}),
the relation between $z$ and $z'$ is given by
\begin{equation}
  \frac{4 (z'^2-t^2)}{\left[ z'^2-(1+t^2) \right]^2}
  = \frac{(1-z^2)^2-4 q^2 z^2}{4 (1+q^2) z^2}
\end{equation}
for $\vev{ B'_t | Q | B'_t }$, and
\begin{equation}
  \frac{4 (z'^2-t^2)}{\left[ z'^2-(1+t^2) \right]^2}
  = \frac{z^2 (z^2-3)^2-q^2 (1-3 z^2)^2}{(1+q^2)(1-3 z^2)^2}
\end{equation}
for $\vev{ B'_t \ast B'_t | B'_t }$.


\renewcommand{\baselinestretch}{0.87}

\begingroup\raggedright\endgroup
\end{document}